\documentclass[%
 reprint,
superscriptaddress,
 amsmath,amssymb,
 aps,
]{revtex4-2}

\usepackage{graphicx}
\usepackage{dcolumn}
\usepackage{hyperref}
\usepackage{textcomp,gensymb}
\usepackage{amsmath,amssymb}
\usepackage{siunitx}

\begin{document}

\preprint{APS/123-QED}

\title{Hard X-ray Generation and Detection of Nanometer-Scale Localized Coherent Acoustic Wave Packets in SrTiO$_3$ and KTaO$_3$}

\author{Yijing Huang}
 \email{huangyj@illinois.edu}
\affiliation{Department of Physics, University of Illinois at Urbana-Champaign, Urbana, IL 61801, USA}
\affiliation{Stanford PULSE Institute, SLAC National Accelerator Laboratory, Menlo Park, CA 94025, USA}
\affiliation{Department of Applied Physics, Stanford University, Stanford, CA 94305, USA}
\affiliation{Stanford Institute for Materials and Energy Sciences, SLAC National Accelerator Laboratory, Menlo Park, CA 94025, USA}

\author{Peihao Sun}%
 \email{peihao.sun@unipd.it}
\affiliation{Stanford PULSE Institute, SLAC National Accelerator Laboratory, Menlo Park, CA 94025, USA}
\affiliation{Dipartimento di Fisica e Astronomia ``Galileo Galilei'', Universit{\`a} degli Studi di Padova, Padova 35131, Italy}

\author{Samuel W. Teitelbaum}
\affiliation{Stanford PULSE Institute, SLAC National Accelerator Laboratory, Menlo Park, CA 94025, USA}
\affiliation{Department of Physics, Arizona State University, Tempe, AZ 85287, USA}

\author{Haoyuan Li}
\author{Yanwen Sun}
\author{Nan Wang}
\author{Sanghoon Song}
\author{Takahiro Sato}
\author{Matthieu Chollet}
\affiliation{Linear Coherent Light Source, SLAC National Accelerator Laboratory, Menlo Park, CA 94025, USA}

\author{Taito Osaka}
\author{Ichiro Inoue}
\affiliation{RIKEN SPring-8 Center, 1-1-1 Kouto, Sayo-cho, Sayo-gun, Hyogo 679-5148, Japan}

\author{Ryan A. Duncan}
\affiliation{Stanford PULSE Institute, SLAC National Accelerator Laboratory, Menlo Park, CA 94025, USA}
\affiliation{Stanford Institute for Materials and Energy Sciences, SLAC National Accelerator Laboratory, Menlo Park, CA 94025, USA}

\author{Hyun D. Shin}
\affiliation{Department of Chemistry, Massachusetts Institute of Technology, Cambridge, MA 02139, USA}

\author{Johann Haber}
\affiliation{Stanford PULSE Institute, SLAC National Accelerator Laboratory, Menlo Park, CA 94025, USA}

\author{Jinjian Zhou}
\author{Marco Bernardi}
\affiliation{Department of Applied Physics and Materials Science, California Institute of Technology, Pasadena, CA 91125, USA}

\author{Mingqiang Gu}
\affiliation{Department of Physics, Southern University of Science and Technology (SUSTech), Shenzhen, Guangdong, China, 518055}

\author{James M. Rondinelli}
\affiliation{Department of Materials Science and Engineering, Northwestern University, Evanston, IL 60208-3108, USA}

\author{Mariano Trigo}
\affiliation{Stanford PULSE Institute, SLAC National Accelerator Laboratory, Menlo Park, CA 94025, USA}
\affiliation{Stanford Institute for Materials and Energy Sciences, SLAC National Accelerator Laboratory, Menlo Park, CA 94025, USA}

\author{Makina Yabashi}
\affiliation{RIKEN SPring-8 Center, 1-1-1 Kouto, Sayo-cho, Sayo-gun, Hyogo 679-5148, Japan}

\author{Alexei A. Maznev}
\author{Keith A. Nelson}
\affiliation{Department of Chemistry, Massachusetts Institute of Technology, Cambridge, MA 02139, USA}

\author{Diling Zhu}
\affiliation{Linear Coherent Light Source, SLAC National Accelerator Laboratory, Menlo Park, CA 94025, USA}

\author{David A. Reis}
 \email{dreis@stanford.edu}
\affiliation{Stanford PULSE Institute, SLAC National Accelerator Laboratory, Menlo Park, CA 94025, USA}
\affiliation{Department of Applied Physics, Stanford University, Stanford, CA 94305, USA}
\affiliation{Stanford Institute for Materials and Energy Sciences, SLAC National Accelerator Laboratory, Menlo Park, CA 94025, USA}

\date{\today}

\begin{abstract}
We demonstrate that the absorption of femtosecond hard x-ray pulses can excite quasi-spherical, high-amplitude and high-wavevector coherent acoustic phonon wavepackets using an all hard-x-ray pump-probe scattering experiment. The time- and momentum-resolved diffuse scattering signal is consistent with strain pulses induced by the rapid electron cascade dynamics following photoionization at uncorrelated excitation centers.
We quantify key parameters of this process, including the localization size of the strain wavepacket and the photon energy conversion efficiency into elastic energy. The parameters are determined by the photoelectron and Auger electron cascade dynamics, as well as the electron-phonon interaction. 
In particular, we obtain the localization size of the observed strain wave packet to be 1.5 and 2.5 nm for bulk SrTiO$_3$ and KTaO$_3$ single crystals, even though there are no nanoscale structures or light-intensity patterns that would ordinarily be required to generate acoustic waves of wavelengths much shorter than the penetration depth.  Whereas in GaAs and GaP we do not observe a signal above background.
The results provide crucial information on the mechanism of x-ray energy deposition into matter and shed light on the shortest collective length scales accessible to coherent acoustic phonon generation using x-ray excitation, facilitating future x-ray study of high-wavevector acoustic phonons and thermal transport at the nanoscale. 
\end{abstract}

\maketitle


\section{Introduction}

Fundamental x-ray-matter interactions are typically dominated by photoionization of core electrons creating highly-excited states that initially decay on the femtosecond time-scale through Auger-Meitner decay and characteristic florescence\cite{xdb}.  The subsequent cascade of secondary excited states involves the inelastic scattering of high-energy electrons and to a lesser extent photons. This creates additional core excited states, and a plethora of both single-particle and collective excitations including electron-hole pairs, plasmons, polarons, and phonons \cite{egerton2011electron} in hard condensed matter systems. This exponentially complex process of secondary interactions serves to either induce or avoid radiation damage depending on how effectively it dissipates the high energy density associated with localized x-ray excitation. 

Thus, it is important to understand experimentally the energy relaxation processes and subsequent structural dynamics following x-ray ionization on the relevant length and time scales.  This is particularly critical for experiments that utilize the high flux and short pulse duration of x-ray free electron lasers (XFELs) to create and/or probe atomic-scale dynamics. Even the most robust materials are not immune to single-shot radiation damage in the focused beam of an XFEL where intensities can be high enough to saturate the photoionization cross-section \cite{YoungNe,Nagler} as well as induce multi-photon K-shell absorption \cite{Doumy,Tamasaku,Ghimire}, and Compton scattering\cite{Fuchs2015}.  In recent x-ray pump, x-ray probe experiments on diamond excited beyond the single-shot damage threshold, the atomic motion appeared frozen for the first 20fs  \cite{inoue2016observation}, while in proteins dense-environment effects have been found to strongly affect local radiation damage induced structural dynamics\cite{Nass}.  It is equally important to understand the structural dynamics induced by x-ray absorption below the single-shot damage threshold.  

Here we present the results of x-ray pump, x-ray probe structural dynamics experiments on the oxide perovskites, SrTiO$_3$ and KTaO$_3$ excited at high densities, but below the multi-shot damage threshold. 
We find that the photoionization leads to the sudden excitation of 3-dimensional (3D) coherent acoustic phonon wavepackets with characteristic wavelengths on the order of single nanometer scale, through analysis of the evolution of diffuse scattering in time and momentum (inducing changes in the signal of over 100\% with moderate pulse fluences).   
We model the strain generation and propagation as due to the in-phase addition of coherent acoustic wavepackets originating from a large collection of nanometer stress centers following localized photoionization events at random uncorrelated sites.  We do not observe signatures of acoustic phonon generation above noise in semiconducting GaAs or GaP indicating that there are significant differences in the cascade process and in particular the dissipation of electronic energy to the lattice.   

 The results have fundamental implications for our understanding of x-ray matter interactions at modest intensities below the damage threshold. In particular, the structural dynamics initiated by the electron cascade process has practical implications for developing a microscopic understanding of condensed matter dynamics, for example, using high wavevector x-ray transient grating spectroscopy\cite{foglia2023extreme} to study nanoscale thermal transport \cite{nanoscale1,nanoscale2}.

\section{Methods}
\begin{figure*}
\centering
\includegraphics[width=0.9\linewidth]{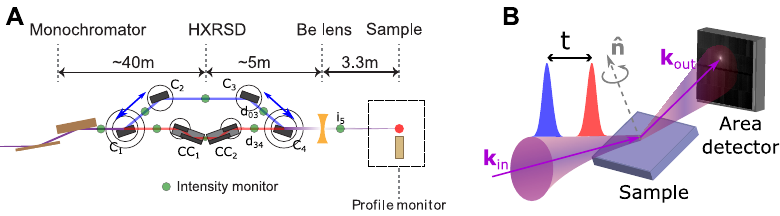}
\caption{
The split-delay setup and experimental geometry. (A) Schematic diagram of the split-delay setup. After the FEL x-ray pulse passes through an upstream double-crystal, diamond (111) monochromator, it arrives at the HRXSD unit and is split into two branches by a silicon crystal with a polished edge (C$_1$): the fixed-delay branch (red lines) consisting of two channel-cut crystals (CC$_1$ and CC$_2$), and the variable-delay branch (blue lines) consisting of four crystals (C$_1$ to C$_4$). Silicon (220) reflections were used for all crystals of the HRXSD unit. X rays from the two branches are combined after crystal C$_4$ and are focused by a Be lens stack onto the sample. The relative delay between the two branches is adjusted by changing the path length in the variable-delay branch, specifically by changing the positions of C$_2$ and C$_3$ using linear translation stages aligned along the blue double-headed arrows. The black circles around the crystals denote the rotation motor stages. Green dots indicate X-ray intensity monitors. (B) Schematic diagram of the experimental geometry. X rays from the two branches, denoted with red and blue pulses, separated by $t$ in time, are focused onto the same position on the sample at an incidence angle of $5\degree$.  The sample is rotated around its surface normal $\mathbf{\hat{n}}$ to go on and off the Bragg condition. The scattered x-rays are collected by an area detector. 
}
\label{fig:setup}
\end{figure*}

The experiment is carried out at the x-ray correlation spectroscopy (XCS) endstation at the Linac Coherent Light Source (LCLS)~\citep{robert_X-ray_2013}. The photon energy is set to 9.828\,keV, slightly below the Ta L3 edge. A schematic diagram of the split-delay setup is shown in Fig.~\ref{fig:setup}A.
The hard x-ray split-delay (HXRSD) unit~\citep{Zhu2017_Development} is inserted into the x-ray beam path, splitting each x-ray pulse into two branches, a fixed-delay branch (red lines) and a variable-delay branch (blue lines). The relative delay between the pulse from the two branches is adjusted  by changing the path length in the variable-delay branch, as indicated by the blue double-headed arrows; in this work, the delay is changed between -2\,ps and 10\,ps in 0.1\,ps steps. After the crystal C$_4$, the pulses from the two branches, each approximately 30\,fs in duration, become nearly collinear and are focused by a beryllium (Be) lens stack of focal length 3.5\,m to approximately $20\,\mathrm{\micro m} \times 20\,\mathrm{\micro m}$ at the sample position. The spatial overlap between the two pulses is optimized with the help of a beam profile monitor consisting of a Ce:YAG scintillator screen positioned in the same plane as the sample and a microscope objective.

Due to imperfections in the translation stages, the angles of crystals C$_2$ and C$_3$ vary slightly as the delays is scanned.  While the magnitude of the angular deviation is small compared to the $\sim16\,\micro$rad Darwin width (for the $p$-polarized x rays), this ``wobble'' nonetheless results in slight variation of the pointing between the two pulses.  
Since the wobbling motion is correlated with the motor positions (which correspond to different delay times), the variations in the pointing are repeatable and thus are partially corrected by changing the angles of crystals C$_2$ and C$_3$ as a function of delay. The remaining variations are well characterized, and the effect on the signal is accounted for using an overlap correction factor as a function of the delay; more details are provided in Appendix~\ref{sec:overlap_correction}.

The pulse energies are measured shot-to-shot at the 120 Hz repetition rate of the FEL by intensity monitors shown as green dots in Fig.~\ref{fig:setup}A. Specifically, the pulse energies in the individual branches are measured by the x-ray diodes d$_{03}$ and d$_{34}$ placed right before the recombination of the branches, while the overall pulse intensity is measured by the intensity monitor i$_5$ placed between the Be lens stack and the sample. The conversion from diode reading to pulse energy is calibrated, as detailed in Appendix~\ref{sec:diode_calibration}.

The experimental geometry is shown in Fig.~\ref{fig:setup}B. The samples are placed in reflection geometry at room temperature, with the beam incident angle on the sample fixed to $5\degree$ grazing. The incident x-ray fluence is kept below the multiple pulse damage threshold of the sample. The x rays scattered by the sample are collected by an area detector (Jungfrau-1M, pixel size $75\,\mathrm{\micro m} \times 75\,\mathrm{\micro m}$)~\cite{Mozzanica2014} placed around 130\,mm away from the sample. In the elastic scattering limit, each pixel on the detector maps to a $\mathbf{Q}=\mathbf{k}_\text{out}-\mathbf{k}_\text{in}$, where $\mathbf{k}_\text{in}$ and $\mathbf{k}_\text{out}$ are the incoming and outgoing wave vectors, respectively, with amplitudes $|\mathbf{k}_\text{in}|=|\mathbf{k}_\text{out}|=2\pi/\lambda$ where $\lambda$ is the x-ray wavelength 1.26\,\AA.
The sample is rotated around its surface normal $\mathbf{\hat{n}}$ until the Bragg condition for a low-order Bragg peak was found, and then rotated by at most $1\degree$ to tune off the Bragg peak to access the diffuse scattering about the peak. For the cubic perovskite samples SrTiO$_{3}$ and KTaO$_{3}$ with surface normal (001), the targeted Bragg peak was $(\bar{1}\bar{1} 2)$.

\section{Results}
\subsection{Extraction of the pump-probe signal}
We begin by examining the general features of the pump-probe signal, taking SrTiO$_{3}$ as an example. The detector measures  x rays from both pulses, such that the scattered intensity detected, 
\begin{equation} \label{eq:I=I0+DeltaI}
    I(\mathbf{Q}, t; \mathcal{E}_1, \mathcal{E}_2) = \mathcal{E}_1 S_0(\mathbf{Q})  + \mathcal{E}_2 S_0(\mathbf{Q}) + \Delta I(\mathbf{Q}, t; \mathcal{E}_1, \mathcal{E}_2),
\end{equation}
where $t$ is the time delay. $\mathcal{E}_1$ and $\mathcal{E}_2$ denote the pulse energies in the variable-delay and fixed-delay branches, respectively, which are measured separately as shown in Fig.~\ref{fig:setup}A. Here and in the rest of the text, $\mathbf{Q}$ denotes the scattering wavevector, $\mathbf{G}$ the nearest reciprocal lattice vector (i.e., the Bragg peak), and $\mathbf{q} \equiv \mathbf{Q}-\mathbf{G}$ the reduced wave vector (i.e., the deviation from the Bragg peak). The first two terms on the right-hand side of Eq.~(\ref{eq:I=I0+DeltaI}) represent the intensities of diffuse scattering in thermal equilibrium, which are proportional to the pulse energies, and $S_0(\mathbf{Q})$ is the diffuse scattering structure factor independent of the pulse energies. The last term, $\Delta I(\mathbf{Q}, t; \mathcal{E}_1, \mathcal{E}_2)$, represents the pump-probe signal which depends on both the pump and probe pulse energies and the relative delay between the two pulses.

\begin{figure*}
\centering
\includegraphics{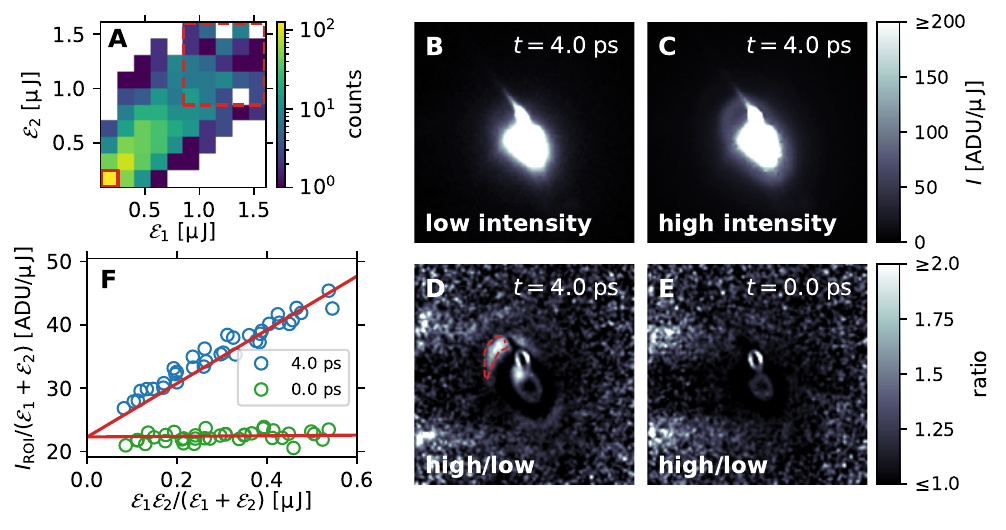}
\caption{
General features of the pump-probe signal. (A) 2D histogram of the distribution of $(\mathcal{E}_1, \mathcal{E}_2)$ at $t=4.0$\,ps; the solid (dashed) box indicates the range corresponding to the low (high) intensity image. (B) and (C) show the normalized low-intensity and high-intensity images at $t=4.0$~ps, whose ratio is shown in (D). (E) shows the ratio at $t=0.0$\,ps following the same procedure, which does not exhibit the ripple-like feature in (D). (F) shows the sum over the ROI indicated by the dashed red line in (D), plotted against the value of $\mathcal{E}_1 \mathcal{E}_2/(\mathcal{E}_1+\mathcal{E}_2)$ at the two delays. Only bins with at least 5 counts are considered. Red lines show linear fits fixing the intercept to be the average value for $t=0.0$\,ps.}
\label{fig:signal_extraction}
\end{figure*}

To extract the pump-probe signal $\Delta I(\mathbf{Q}, t; \mathcal{E}_1, \mathcal{E}_2)$, we first note that the x-ray pulse intensity delivered onto the sample varies shot-to-shot due to the fluctuating overlap between the x-ray spectrum coming into the split-and-delay system and the band-pass of the crystals in the system~\cite{zhu2014performance}. 
The ratio between the intensities in the two branches, $\mathcal{E}_1/\mathcal{E}_2$, also fluctuates due to jitter in the beam position at the splitting crystal C$_1$. Therefore, throughout the measurement, we collect a large set of images with a wide distribution of pulse energies $\mathcal{E}_1$ and $\mathcal{E}_2$. As an example, a histogram of the distribution of $(\mathcal{E}_1, \mathcal{E}_2)$ at delay $t=4.0$\,ps is shown in Fig.~\ref{fig:signal_extraction}A. The distributions at other time delays are similar. This wide distribution of $(\mathcal{E}_1, \mathcal{E}_2)$ helps isolate the pump-probe signal $\Delta I(\mathbf{Q}, t; \mathcal{E}_1, \mathcal{E}_2)$: from all shots at time delay $t$, we select ``low intensity'' ones ($0.1\,\text{\micro J} < \mathcal{E}_1, \mathcal{E}_2 < 0.25\,\text{\micro J}$) where the pump-probe signal is expected to be small, and ``high intensity'' ones ($0.85\,\text{\micro J} < \mathcal{E}_1, \mathcal{E}_2 < 1.6\,\text{\micro J}$) where the pump-probe signal should be large. These ranges are indicated by the solid and dashed boxes in the histogram in Fig.~\ref{fig:signal_extraction}A. We then calculate the normalized image for each category by dividing the summed image by the summed pulse intensities.

The normalized low- and high-intensity images for SrTiO$_{3}$ at $t=4.0$\,ps are shown in Fig.~\ref{fig:signal_extraction}B-C. Note that the long white streaks are due to scattering from the tails of the Bragg peak (from the surface truncation rod). Comparing these two images, one can see modulations away from the central region appear in the high-intensity image, which becomes clearer when dividing the high-intensity image by the low-intensity one as shown in Fig.~\ref{fig:signal_extraction}D. These modulations appear like ripples emanating from the center, which corresponds to the closest point to the $(\bar{1}\bar{1}2)$ Bragg peak on the detector (i.e., on the Ewald sphere), reflecting the acoustic phonon excitation in the sample. Note that the relative signal level is rather high: the modulations reach more than 100\% of the diffuse scattering background approximated by the low-intensity image in Fig.~\ref{fig:signal_extraction}B. In comparison, the pump-probe signal appears negligible around zero time delay: Fig.~\ref{fig:signal_extraction}E shows the results for delay $t=0.0$\,ps, which does not contain any modulation like in Fig.~\ref{fig:signal_extraction}D. Therefore, we use the data at time zero as the background diffuse scattering, as will be further detailed below.

Having observed the general features of the pump-probe signal, we next demonstrate that it is bi-linear in the pump and probe pulse energies. Because the pump-probe signal, $\Delta I(\mathbf{Q}, t; \mathcal{E}_1, \mathcal{E}_2)$, should be proportional to both the probe pulse energy and the amount of lattice distortion created by the pump pulse, the bi-linearity is expected if the latter is proportional to the number of photons in the pump. In this case, we may write $\Delta I(\mathbf{Q}, t; \mathcal{E}_1, \mathcal{E}_2) = C(\mathbf{Q},t)\mathcal{E}_1 \mathcal{E}_2$, where $C(\mathbf{Q},t)$ is the pump-probe response coefficient independent of the pulse energies. In this case, the normalized scattered intensity, 
\begin{equation}
    \frac{I(\mathbf{Q}, t; \mathcal{E}_1, \mathcal{E}_2)}{\mathcal{E}_1+\mathcal{E}_2} = S_0(\mathbf{Q}) + C(\mathbf{Q},t) \frac{\mathcal{E}_1 \mathcal{E}_2}{\mathcal{E}_1+\mathcal{E}_2}. \label{eq:non-linear-signal}
\end{equation}

We now test the validity of Eq.~(\ref{eq:non-linear-signal}). Using the extracted pump-probe signal in Fig.~\ref{fig:signal_extraction}D, we select a region of interest (ROI) with a clear signal, as indicated by the red dashed line. Fig.~\ref{fig:signal_extraction}F  shows the summed intensity within this region, $I_\text{ROI}$, normalized by the total pulse energy $\mathcal{E}_1+\mathcal{E}_2$, plotted as a function of $\mathcal{E}_1 \mathcal{E}_2/(\mathcal{E}_1+\mathcal{E}_2)$. The results for delay $t=4.0$\,ps and 0.0\,ps are shown as blue and green circles, respectively, where each circle corresponds to the average over a bin in the histogram in Fig.~\ref{fig:signal_extraction}A with at least 5 shots. These data are consistent with a linear trend with the same intercept at $\mathcal{E}_1\mathcal{E}_2=0$, which supports the validity of Eq.~(\ref{eq:non-linear-signal}) and hence the bi-linearity of the pump-probe signal. Therefore, the results verify our expectation that the total lattice distortion is proportional to the pump pulse energy. Moreover, while the data for $t=4.0$\,ps shows a clear slope, the data for $t=0.0$ appear independent from the pulse energies, confirming the absence of pump-probe signal at zero time delay.

\begin{figure*}
    \centering
    \includegraphics
    {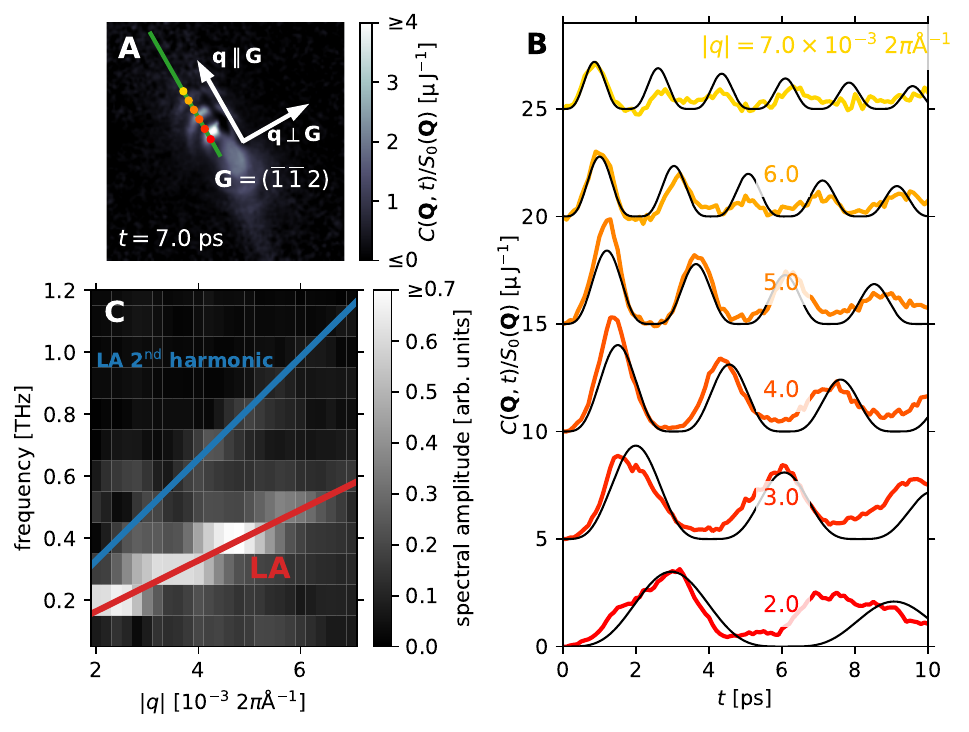}
    \caption{
    Measured x-ray pump, x-ray probe signal $C(\mathbf{Q},t)/S_0(\mathbf{Q})$ in SrTiO$_{3}$.  
    (A) $C(\mathbf{Q},t)/S_0(\mathbf{Q})$ at $t=7.0$ ps. The green line shows the direction $\mathbf{Q}\parallel\mathbf{G}$, which coincides with the direction of the largest intensity modulation. (B) The time dependence of $C(\mathbf{Q},t)/S_0(\mathbf{Q})$ at selected wavevectors $\mathbf{q}$ along the red line in (A). The corresponding locations on the detector are indicated as colored dots in (A). An offset is added between traces of different $|q|$ values for clarity; $C(\mathbf{Q},t)/S_0(\mathbf{Q})$ is zero at $t=0$. The black lines are fit results, to be discussed in the ``Model'' section. (C) Fourier transform spectral amplitudes along the direction of the red line in (A). The red and blue lines show the dispersion of the LA phonon and the LA second harmonic obtained from DFT calculations.} 
    \label{fig:time_traces}
\end{figure*}

Since we have demonstrated that the pump-probe signal is negligible around zero delay, to increase the signal-to-noise ratio, we use the normalized intensity including all valid shots at $t=0.0$\,ps, $I^\text{norm}(\mathbf{Q}, t=0)$, as the equilibrium diffuse scattering structure factor $S_0(\mathbf{Q})$, in the absence of the effect of the pump. Using Eq.~(\ref{eq:non-linear-signal}), the pump-probe coefficient at delay $t$ is thus be obtained from the experimental data set as:
\begin{equation} \label{eq:pump_probe_coefficient}
    \frac{C(\mathbf{Q},t)}{S_0(\mathbf{Q})} = \left[ \frac{I^{\text{norm}} (\mathbf{Q}, t)}{I^{\text{norm}} (\mathbf{Q}, t=0)} - 1 \right] \frac{\sum_s (\mathcal{E}_1^{(s)} + \mathcal{E}_2^{(s)})}{\sum_s \mathcal{E}_1^{(s)}\mathcal{E}_2^{(s)}} [\mathcal{O}(t)]^{-1},
\end{equation}
where the sum is over all shots $s$ at delay $t$. Here, $\mathcal{O}(t)$ denotes the correction factor of order unity which accounts for changes in the overlap between the two beams on the sample during the delay scan due to the aforementioned wobbling motion of the delay scan stages (see Appendix~\ref{sec:overlap_correction}).

An example of the pump-probe signal, obtained using Eq.~(\ref{eq:pump_probe_coefficient}) for $t=7.0$\,ps, is shown in Fig.~\ref{fig:time_traces}A. The green line shows the direction $\mathbf{q}\parallel\mathbf{G}$, which coincides with the direction of the largest intensity modulation. Along this line, we take several $\mathbf{q}$ points (indicated by the colored dots) and plot the time dependence of the pump-probe signal in Fig.~\ref{fig:time_traces}B, where the labels indicate the magnitude $q \equiv |\mathbf{q}|$ for each trace. These curves exhibit damped oscillations, whose frequency increases with increasing $q$. The curves do not resemble a perfect sinusoidal function but feature flat minima, suggesting the existence of even-order frequency overtones. With a Fourier transformation, we obtain the spectral weights along this $\mathbf{q}$ direction, which are shown in Fig.~\ref{fig:time_traces}C.

The results indicate that the excited modes are predominantly LA phonons. Firstly, the direction of the strongest modulation (green line in Fig.~\ref{fig:time_traces}A) coincides with the direction $\mathbf{q}\parallel\mathbf{G}$, while the modulation vanishes in the perpendicular direction, consistent with the $|\mathbf{Q}\cdot\mathbf{\epsilon}|^2$ dependence in the scattering intensity where $\mathbf{\epsilon}$ is the phonon polarization vector. Secondly, we overlay the spectral weights in Fig.~\ref{fig:time_traces}C with the LA phonon dispersion in the direction $\mathbf{q}\parallel\mathbf{G}$ [using $v=8.2$\,km/s, which is calculated by density-functional theory (DFT) along the selected $\mathbf{q}$ direction] and its second-harmonic overtone, showing good agreement with the data.

\subsection{Model}
We present a model that is consistent with our observations and describes quantitatively the time evolution of the pump-probe signal.
The model is based on the following physical picture. First, the stochastic absorption of x-ray photons from the pump pulse causes the creation of a large number of uncorrelated photoelectrons and core holes, each of which relaxes generating a cascade of lower-energy electrons. This process is mostly complete within 100\,fs~\cite{Ziaja2005}, much faster than the period of the acoustic phonons that we detect. After this process, a large number of electron clouds are formed within the sample. These clouds are expected to have a core region, on the order of several nanometers, with a high electron density~\cite{Ziaja2005,gibrekhterman1993spatial}, which serves as a random collection of excitation centers. The high concentration of secondary photoelectrons about each center leads to a sudden local stress that produces a propagating strain pulse in the form of a coherent longitudinal acoustic phonon wavepacket with a typical phonon period given by the time it takes for sound to propagate across the core region of the cascade. 

The probability of absorption about any given atomic site is much less than one and is given by the product of the photon fluence and the photoionization cross-section. For SrTiO$_3$ at 9.828 keV, it is dominated by absorption on the Sr sites with a mean distance between absorption events on order of 30\,nm, for 0.5 $\mu$J in a $20\,\mathrm{\micro m} \times 20\,\mathrm{\micro m}$ spot. 
This is an order of magnitude larger than the inverse of the maximum $q$ in Fig.~\ref{fig:time_traces}A with observable ``ripple'' feature, which is around $1/(5\times 10^{-3}\,2\pi\mathrm{\AA}^{-1}) \approx 3\,\mathrm{nm}$. Therefore, we assume that the interference between strain waves from the individual random photoabsorption events largely averages out. Furthermore, since x-ray photoabsorption is a stochastic process, we assume that the spatial distribution of these excitation centers across the different unit cells is given by a binomial probability distribution. 

Since we measure the incoherent sum of their scattering amplitudes (more details below and in Appendix~\ref{sec:model_derivations}), it is justified to take the ensemble average limit when describing the strain generation and propagation. Although individual photoelectrons may create anisotropic distributions of secondary electrons~\cite{gibrekhterman1993spatial}, it is expected to become small by 100\,fs~\cite{Ziaja2005}, and the distribution is assumed to be isotropic in the ensemble average limit~\cite{gibrekhterman1993spatial}.

Taking into account the arguments above, we build a model assuming that: 
1) The pump pulse creates a number of excitation centers that are randomly and sparsely distributed within the illuminated volume, and the number of these centers is proportional to the pump fluence.
2) Around each excitation center, a step-function-like (in time) stress field causes a sudden change in the equilibrium lattice constant and therefore a sudden strain. We assume the excitation is instantaneous compared to the phonon periods which are on the order of picoseconds (see Fig.~\ref{fig:time_traces}C), so at $t=0$ the atomic displacements are zero.
3) The strain field is isotropic and assumes a Gaussian spatial profile in the ensemble average limit.
4) The strain field can be treated in the continuum limit, since the smallest length scales considered (several nanometers, corresponding to the inverse of the maximum $q$ range of visible ripples) are still significantly larger than the size of the unit cell.  Furthurmore for simplicitly, we approximate the material as elastically isotropic. Under these assumptions, the Fourier transform of the average displacement field for a single excitation center is (see Appendix~\ref{sec:model_derivations} for detailed derivations):
\begin{equation}
    \Tilde{\mathbf{u}}(\mathbf{q}, t) = \frac{i\pi^{3/2} A \sigma^2}{V q} e^{-\sigma^2 q^2 / 4} \left[1 - \cos\left(q v t \right) \right]e^{-t/\tau} \hat{\mathbf{q}}, \label{eq:u(q,t)_main}
\end{equation}
where $A$ describes the amplitude of the displacement field, $\sigma$ is the rms extent of the distortion field, $v=8.2$\,km/s is the velocity of the LA wave obtained from data in Fig~\ref{fig:time_traces}C, $e^{-t/\tau}$ is a phenomonological decay decay factor added to account for the observed decay of the oscillations (see Fig.~\ref{fig:time_traces}B), and $\hat{\mathbf{q}}$ is the unit vector in the direction of $\mathbf{q}$. Here, $\Tilde{\mathbf{u}}(\mathbf{q},t)$ has the unit of length.
The $\left[1 - \cos\left(q v t \right) \right]$ term is typical of displacive-like excitation, where the equilibrium position of the lattice suddenly shifts and atoms oscillate around the newequilibrium~\cite{zeiger1992theory}.  We take a common decay time $\tau$, for both the decay of the new equilibrium back to the original equilibrium, and the oscillation amplitude.

Since we observe that the modulations of the diffuse scattering (see Fig.~\ref{fig:time_traces}) happen at the regime of relatively small $q \equiv |\mathbf{q}| \ll |\mathbf{G}|$, and we assume that the spatial distribution of excitation centers is sparse and random, the change in diffuse scattering intensity due to the distortions is derived as for the Huang diffuse scattering due to static defects~\cite{dederichs_theory_1973}. Hence the intensity modulation, 
\begin{equation} \label{eq:diffuse_intensity_modulation}
    \Delta I(\mathbf{Q},t) \propto c |\mathbf{G} \cdot \Tilde{\mathbf{u}}(\mathbf{q},t)|^2 \mathcal{E}_1,
\end{equation}
where $\mathcal{E}_1$ is the probe pulse energy; $c \ll 1$ is the concentration (number per unit cell) of excitation centers that are expected to be proportional to the pump pulse energy $\mathcal{E}_2$. Thus, $\Delta I(\mathbf{Q},t)$ is proportional to $\mathcal{E}_1 \mathcal{E}_2$ as expected. The full expression for $\Delta I(\mathbf{Q},t)$ considering all geometric factors is provided in SI. Note that in Eq.~(\ref{eq:diffuse_intensity_modulation}), the term $|\mathbf{G} \cdot \Tilde{\mathbf{u}}(\mathbf{q},t)|^2$ gives rise to the angular dependence $\Delta I(\mathbf{Q},t) \propto |\mathbf{G} \cdot \hat{\mathbf{q}}|^2$, in agreement with the experimental observation in Fig.~\ref{fig:time_traces}, even for an isotropic  $\Tilde{\mathbf{u}}(\mathbf{q},t) = \Tilde{\mathbf{u}}(|q|,t)$. The thermal-equilibrium diffuse scattering $I_0(\mathbf{Q}, t)$, on the other hand, is presumed to be dominated by thermal phonons, for simplicity. The expression for thermal diffuse scattering is given in Eq.~(\ref{eq:TDS_1st_order}). 

Based on this model, the pump-probe signal is (see Appendix~\ref{sec:model_derivations} for detailed derivations), 
\begin{equation} \label{eq:pump_probe_signal_model}
    \frac{C(\mathbf{Q},t)}{S_0(\mathbf{Q})} = \mathcal{F} \sigma^3 \left( \frac{U_p}{U_d} \right) e^{-\frac{\sigma^2 q^2}{2}} \left[1 - \cos\left(q v t \right) \right]^2 e^{-\frac{2 t}{\tau}} \left|\mathbf{G} \cdot \hat{\mathbf{q}} \right|^2,
\end{equation}
where the pre-factor $\mathcal{F}$ takes into account (see Eq.~(\ref{eq:pre_factor_F}) for the full expression): geometric factors (e.g., the beam size), the x-ray linear absorption coefficient, thermal diffuse scattering background assuming phonon frequencies and eigenvectors as obtained from DFT, as well as other known constants (e.g., x-ray atomic scattering form factors at the given $q$ and photon energy), all of which are independent of parameters of the model. Thus, the pre-factor $\mathcal{F}$ can be calculated for any given $\mathbf{Q}$. We only explicitly write out in Eq.~(\ref{eq:pump_probe_signal_model}) the following terms: the time dependence $\left[1 - \cos\left(q v t \right) \right]^2 e^{-2 t/\tau}$, the angular dependence $\left|\mathbf{G} \cdot \hat{\mathbf{q}} \right|^2$ (which determines the intensity anisotropy of the ``ripples'' in Figure~\ref{fig:time_traces}A), the size of the distortion field $\sigma$, and the energy conversion coefficient $U_p/U_d$. Here $U_d$ is the absorbed energy density and $U_p$ is the energy density of the launched acoustic phonons, both defined in the bulk average limit.

\begin{figure*}
    \centering
    \includegraphics
    {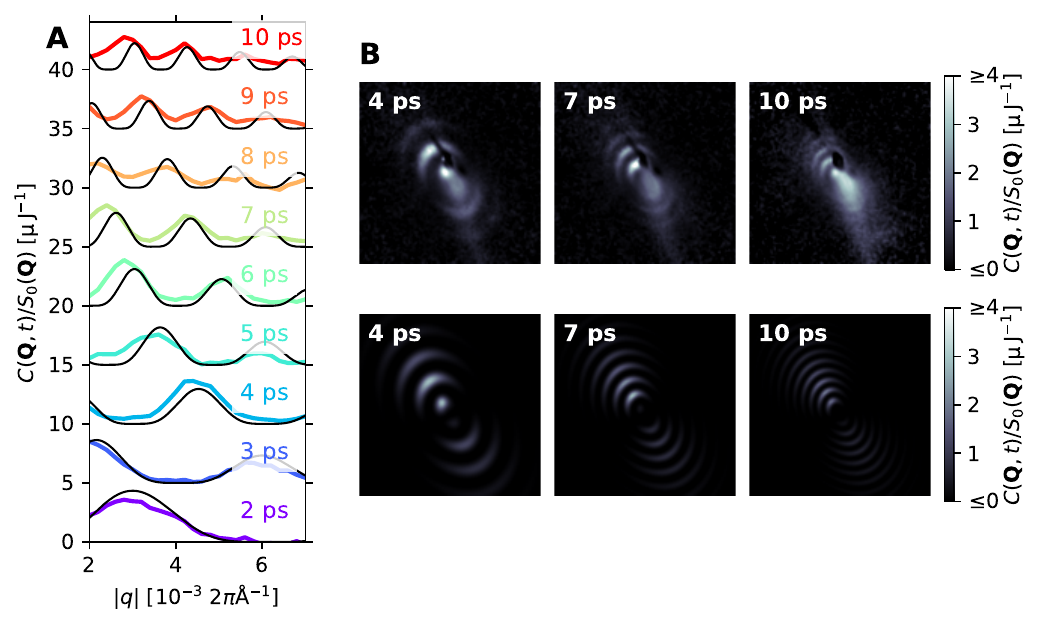}
    \caption{
    Time-dependence of the pump-probe signal in SrTiO$_{3}$, and its model predictions using fit parameters.  
    (A)  The pump-probe signal $C(\mathbf{Q},t)/S_0(\mathbf{Q})$ along the $\mathbf{q}$ linecut in Fig.~\ref{fig:time_traces}A, at selected delay times. Colored lines show the experimental data, while black lines are predictions by the model.
    (B) $C(\mathbf{Q},t)/S_0(\mathbf{Q})$ on the area detector compared between the experiment (top row) and the model predictions (bottom row) at $t=4, 7, 10$\,ps.
    } 
    \label{fig:fit_results}
\end{figure*}

Using Eq.~(\ref{eq:pump_probe_signal_model}), we fit our model to the experimentally measured $C(\mathbf{Q}, t)/S_0(\mathbf{Q})$ to extract the main physical quantities of interest: the size of the distortion field, $\sigma$, and the energy conversion efficiency, $U_p/U_d$. 
The fit is done in the following way: first, we estimate the decay constant $\tau$ with the time-dependent $C(\mathbf{Q}, t)/S_0(\mathbf{Q})$, the colorful traces in Fig.~\ref{fig:time_traces}B, assuming that $\tau$ is independent of $\mathbf{q}$ (i.e., a global estimate to all traces in Fig.~\ref{fig:time_traces}B).
Then, we vary the parameters $\sigma$ and $U_p/U_d$ to best fit the model to the data in the $q$-range from 2 to $7\times 10^{-3}\;2\pi\mathrm{\AA}^{-1}$ along the direction $\mathbf{q}\parallel\mathbf{G}$ (i.e., the green line cut in Fig.~\ref{fig:time_traces}A) and in the available delay range from 0 to 10\,ps. Data at $q>7\times 10^{-3}\; 2\pi\mathrm{\AA}^{-1}$ are excluded because of low signal levels, while data at $q<2 \times 10^{-3}\;2\pi \mathrm{\AA}^{-1}$ are excluded because of their sensitivity to inaccuracies in $\mathbf{q}$-space calibration and in the modeling of the diffuse scattering, which may contain a background from static disorder besides the thermal diffuse scattering considered above. The results are presented in Fig.~\ref{fig:fit_results}, which shows the measured pump-probe signal $C(\mathbf{Q},t)/S_0(\mathbf{Q})$ (colored lines) and fit results (black lines) as a function of $q$ at different delays. 
The extracted fit parameters are $\sigma=1.5$\,nm, $U_p/U_d = 7 \times 10^{-3}$, and $\tau=12$\,ps for SrTiO$_3$.
The fits are shown as black lines in Fig.~\ref{fig:time_traces} A. 
Fig.~\ref{fig:fit_results}B shows $C(\mathbf{Q}, t)/S_0(\mathbf{Q})$ data on the selected area of the detector (top row) and model predictions using the fit parameters (bottom row), at delays of 4, 7, and 10\,ps. 
Based on the general agreement between the model predictions and the experimental data, we consider our few-parameter model to be robust.
Note, however, that $\sigma$ is model-dependent, and it may change if one assumes a different form of the source profile other than a Gaussian one (e.g., an exponential decay profile in real space).
In KTaO$_3$, the extracted fit parameters are $\sigma=2.5$\,nm and $U_p/U_d = 2\times 10^{-3}$; more detailed results are shown in Appendix~\ref{sec:additional_data}.

\section{Discussion}
It is remarkable that our simple model, which only assumes that spherical strain waves are launched from random, uncorrelated, and three-dimensionally localized sources of electrons, reproduces our experimental data and allows for the quantification of key parameters of this process, including the localization size $\sigma$ of strain wave packets, and the photon energy conversion efficiency $U_p/U_d$ into the elastic waves. 
The model is in stark contrast to ultrafast optical excitation in opaque materials where the energy absorbed is distributed uniformly over the illuminated area and exponentially along a distance (of absorption length) much shorter than the wavelength and the beam size that typically leads to an effective 1D strain wave propagating into the bulk with characteristic wavelength given by the penetration depth.
In the all-x-ray experiment reported here, coherent acoustic phonons propagate in 3D. Even though the absorption of the x-ray on average leads to an exponentially decaying density profile into the bulk, the typical wavelength of coherent acoustic phonons is many orders of magnitude shorter than both the x-ray spot size and the penetration depth, pointing to the fact that X-ray excitation induces much more localized electron distribution than optical pump and the potentially dramatically different electron-phonon coupling mechanism. The generation and detection of coherent high-wavevector acoustic waves as reported here do not involve engineered interfaces or inhomogeneities, such as a transducer layer~\cite{Tas1994,thomsen_surface_1986,henighan2016generation} or a superlattice structure~\cite{Hawker2000,chou2019,trigo_acoustic}, which would normally be required for generation and detection of high-wavevector acoustic waves using optical pulses.

In the case of  SrTiO$_3$, we find $\sigma = 1.5\;$ nm and $U_p/U_d = 7\times 10^{-3}$. As pointed out in the Model section, from $U_p/U_d$ one can obtain the product of the strain amplitude and concentration of localized excitation centers $cA^2$.
If we assume the concentration of excitation sites is equal to the initial density of photoexcited atoms ($\sim 10^{17}\text{ cm}^{-3}$), the amplitude is 0.15\; nm corresponding to a dilation at the excitation center of $\sim$10\%. 
Notably, while we find similar results for the oxide perovskites, SrTiO$_3$ and KTaO$_3$, we do not detect an observable signal for the tetrahedral semiconductors GaAs and GaP over a similar $q$-range. 
We expect the effective source sizes to be similar for the materials if they were solely based on the dependence of the cascade-electron distributions on the atomic constituents~\cite{lipp2022quantifying}. Moreover, the concentrations of initial ionization sites should also be similar based on the photoelectron cross-sections. Thus, we estimate an upper limit for the strain amplitude to be about 30 times smaller than for oxides.  

The dramatic difference in the response between these materials depends on the microscopic details of the strain generation and how it depends on the complex dynamics of the relaxation of the highly excited states and how it couples to the lattice.  In the optical regime, ultrafast excitation of low-energy electrons (and holes) in opaque materials leads to coherent strain generation through both thermoelastic and deformation potential mechanisms.  If a similar process were to dominate the x-ray case, the differences in the material's properties would also not be sufficient to explain the differences.  
However, the detailed spatio-temporal profile of the stress and resultant strain fields depends not just on the thermal expansion coefficient and deformation potentials, but also on the electron cooling rate and whether there is significant transport across the initial excitation region during the sound propagation time\cite{thomsen_surface_1986}.  Even in the optical regime, this can reshape the coherent acoustic phonon pulse enhancing the lower frequency components and suppressing the higher ones, as seen for example in x-ray diffraction experiments from photoexcited Ge\cite{DeCampPRL2003, DeCampJSR2005}.  In the x-ray regime, the length scales are much smaller, and the electron energies are initially much higher, such that the details of the energy deposition rates and in particular the coupling to plasmons and polarons could become important given the high polarizability of the oxides.  In particular, polarons feature local electron-lattice interactions that may explain the high-wave-vector excitation of coherent strain waves~\cite{polaron1,polaron2,polaron3}. 

The average size of the strain field around each excitation center is linked with the distribution of secondary electrons and their coupling to the lattice. The spatial distribution is determined by the energy and momentum relaxation channels of the photoelectrons and Auger electrons~\cite{akkerman1980monte,akkerman1985comparison,grum2017size,gibrekhterman1993spatial,lipp2022quantifying}.
For reference, in SrTiO$_3$ where Sr dominates the photoabsorption, a photoelectron ionized from Sr 2$s$ is $\sim$ 7.6~keV, while the subsequent Auger electron is $\sim$ 1.6~keV~\cite{sasaki1996spectator}.
The secondary electron cascade initiated by a multi-keV electron is expected on average to have a size on the order of hundreds of nanometers. Lower energy electrons, i.e., $<1$ keV, are expected to initiate cascades that end up with a localized secondary electron distribution with a characteristic size on the order of nanometers and a high peak density near the excitation center~\cite{grum2017size,gibrekhterman1993spatial,lipp2022quantifying,Ziaja2005,merlet2012uncertainty}.
The former, while possessing an extended overall dimension, features more localized centers of a few nanometers in size~\cite{gibrekhterman1993spatial}.
The latter has a general length scale consistent with our experimentally measured $\sigma$.
Therefore, given the inelastic mean free path of multi-keV electrons is on the order of tens of nanometers~\cite{Stein1976,Salvat1985,Reich1988,akkerman2009monte}, much longer than the experimentally measured $\sigma$, implies that $c$ exceeds the initial excitation density, and thus our estimate for $A$ is an upper one. 
The spherical wave packet center concentration $c$ can indeed be lower than solely determined by the material photoabsorption cross-section, due to Auger electrons from multiple elements (e.g., both Ti and Sr atoms in SrTiO$_3$), re-absorption of fluorescence photons, and ionization by secondary electrons.

We note that coherent phonons can be selectively generated with light by spatial patterning of the radiation. One such case is the transient grating (TG) technique where two crossed laser pulses create a standing wave interference pattern that excites phonons with the same period.
The TG technique has recently been extended from the optical to extreme ultraviolet (EUV) wavelengths~\cite{PhysRev.133.A759,PhysRevLett.62.1376,Svetina2019,Rouxel2021, bencivenga_four-wave_2015, bencivenga_four-wave-mixing_2016, bencivenga2019nanoscale,foglia2023extreme, EUVTG2023} and has been able to selectively excite phonons with wavelength as small as 24\,nm~\cite{foglia2023extreme}. 
With hard x-ray laser pulses, the period of the standing waves can be reduced to well below the sub-1\,nm scales due to the short x-ray photon wavelength\cite{PhysRev.133.A759,bedzyk2002x}. It has been suggested that the fundamental limit of the wavelength of coherent acoustic phonon generated by such gratings is determined by the inelastic mean free path of electrons~\cite{de2019electron,EUVTG2023} leading to significant signal degradation in the sub-10\,nm length scales.  
The results here show that during the electronic cascade process, significant phonon generation can occur at nanometer length scales before the electronic and thermal excitation homogenizes.

\section{Conclusion}

In summary, we report hard x-ray generation and detection of high-wavevector, large amplitude coherent acoustic strain pulses in oxide insulators. 
We anticipate future experiments with higher signal sensitivity and $q$ resolution to definitively clarify the speculations above.
The key is to directly extract $\sigma$ and $A$ in X-ray-pumped semiconductors.
If indeed $\sigma$ is confirmed to be of similar magnitude as SrTiO$_3$, it will support the mechanism of direct local electron-phonon coupling.
On the other hand, if $\sigma$ turns out to be much larger compared with SrTiO$_3$, it will prompt a more detailed look at the electronic cascade and diffusion process.
Though the electron cascade upon hard x-ray photoabsorption is relatively well understood based on simulations~\cite{akkerman1980monte,akkerman1985comparison,grum2017size,gibrekhterman1993spatial,lipp2022quantifying,
grum2017size,lipp2022quantifying,Ziaja2005,merlet2012uncertainty}, additional simulations of the electron-phonon coupling together with the electron cascade process after photoabsorption of hard x-ray photon will greatly help in understanding the full process of high $q$ coherent phonon generation.

The spectral content of the coherent acoustic phonons that make up the strain wave is consistent with a large collection of localized sources of sudden stress with size on the order of a few nanometers. 
The size is expected to be determined by the complex dynamics of the high-energy electron cascade and is significantly shorter than the x-ray penetration depth. 
The observed excitation site dimension of 1.5\,nm (in SrTiO$_3$) is significantly shorter than the low-energy electron inelastic mean free path~\cite{Stein1976,Salvat1985,Reich1988,akkerman2009monte,de2019electron}.
While a more systematic study is required to determine the excitation mechanism of phonons from the x-ray-induced charge distribution, 
the generation of high amplitude coherent phonon wavepackets with nm-scale characteristic extent substantiates that high amplitude monochromatic acoustic phonons can be generated with sub-10\,nm scale wavelengths using x-ray transient gratings methods, addressing an important length scale for the thermal transport in modern integrated circuits and its power management.

The fraction of x-ray energy deposited in acoustic waves, $U_p/U_d$ on the order of $\sim 10^{-3}$ (We obtain that the energy conversion efficiency $U_p/U_d \sim 7 \times 10^{-3}$ for SrTiO$_3$, and $\sim 2 \times 10^{-3}$ for KTaO$_3$), as obtained from our model, may help quantify an energy transfer channel relevant to radiation damage processes relevant to all FEL based pump-probe measurements for condensed matter physics.
Besides crystalline materials, the reported methods will also be beneficial for studying the x-ray-induced structural changes in amorphous materials on short times scales~\cite{Ruta2017,Dallari2023,Martinelli2023}. 

\begin{acknowledgments}
The authors thank J.~B.~Hastings for useful discussions.
This work was supported by the U.S. Department of Energy, Office of Science, Office of Basic Energy Sciences through the Division of Materials
Sciences and Engineering under Contract No.~DE-AC02-76SF00515. Measurements were carried out at the Linac Coherent Light Source, a national user facility operated by Stanford University on behalf of the U.S. Department
of Energy, Office of Basic Energy Sciences under Contract No. DE AC02- 76SF00515. 
Preliminary experiments were performed at SACLA with the approval of the Japan Synchrotron Radiation Research Institute (JASRI) (Proposal No.2017B8046) 
P.~Sun acknowledges funding from the European Union's Horizon 2020 research and innovation programme under the Marie Sk{\l}odowska-Curie grant agreement No.~101023787. 
The participants from MIT were supported by the Department of Energy, Office of Science, Office of Basic Energy Sciences, under Award Number DE-SC0019126. 
M.G. and J.M.R. were supported by the U.S. Department of Energy (DOE) under Grant No. DE-SC0012375.
\end{acknowledgments}

Y.H., P.S., and S.W.T. contributed equally to this work.

\appendix

\section{\label{sec:model_derivations}Derivations for the model}
\subsection{Spherical wave solution}
In this section, we present the derivation of the spherical strain wave model which is used in the main text. Two main assumptions are made. Firstly, we take the continuum limit, which is appropriate given that we are considering lengths scales of tens of nanometers and above, which is large compared with the size of the unit cell. Secondly, we assume that the material is isotropic, which greatly simplifies the mathematical form of the results. The second assumption is not strictly true in reality, but the analysis and final results are not significantly influenced by the anisotropy of the materials, so we keep this assumption. Furthermore, we start the derivation without considering dissipation, to demonstrate the main features of the propagating spherical waves (the oscillating patterns in reciprocal space that are observed in our data). At the end of the section, we take into account the effects that lead to decay over time.

In our model, an X-ray photon excitation event leads to a distortion in the equilibrium position at time $t=0$. This distortion is assumed to be spherically symmetric, and it launches longitudinal spherical waves for $t > 0$. Since the material is assumed to be isotropic, the spherical symmetry is preserved during the wave propagation; in other words, the displacement field in the material after the excitation, $\mathbf{u}(\mathbf{r}, t)$, should be curl-free. Therefore, we may write $\mathbf{u}(\mathbf{r}, t)=\nabla \phi(\mathbf{r}, t)$, where $\phi(\mathbf{r}, t)$ is a scalar field which satisfies the wave equation~\cite{Chandrasekharaiah1994}:
\begin{equation} \label{eq:general_wave_equation_real_space}
    \nabla^2 \phi(\mathbf{r}, t) - \frac{1}{v^2} \frac{\partial^2}{\partial t^2} \phi(\mathbf{r}, t) = s(\mathbf{r})H(t),
\end{equation}
where $v$ is the longitudinal sound speed and $H(t)$ is the Heaviside step function. $s(\mathbf{r})$ represents the distortion field of the new equilibrium, whose form is not specified at this point. In reciprocal space, Eq.~(\ref{eq:general_wave_equation_real_space}) becomes
\begin{equation} \label{eq:general_wave_equation_reciprocal_space}
    - \left( q^2 + \frac{1}{v^2} \frac{\partial^2}{\partial t^2} \right) \Tilde{\phi}(\mathbf{q}, t) = \Tilde{s}(\mathbf{q})H(t).
\end{equation}

Eq.~(\ref{eq:general_wave_equation_reciprocal_space}) is a standard wave equation whose general solution for $t>0$ is
\begin{equation} \label{eq:wave_equation_general_solution}
    \Tilde{\phi}(q, t) = \Tilde{\phi}_0(q) + F(q) e^{i q v t} + G(q) e^{-i q v t},
\end{equation}
where $\Tilde{\phi}_0(q)=-\Tilde{s}(q)/q^2$ is the equilibrium solution, and $F(q)$ and $G(q)$ are arbitrary functions of $q$. Note that we have now dropped the dependence on the direction of $\mathbf{q}$ because of spherical symmetry.

Now we impose the initial conditions that, at $t=0$, there is no displacement or movement of the atoms:
\begin{align}
    \Tilde{\phi}(q, t=0) =& 0, \label{eq:initial_condition_1}\\
    \left. \frac{\partial \Tilde{\phi}(q, t)}{\partial t} \right|_{t=0} =& 0. \label{eq:initial_condition_2}
\end{align}
This leads to $F(q)=G(q)=-\Tilde{\phi}_0(q)/2$. Hence, the solution, Eq.~(\ref{eq:wave_equation_general_solution}), becomes
\begin{equation} \label{eq:general_solution_inhom_wave_eq}
    \Tilde{\phi}(q, t) = \Tilde{\phi}_0(q) [1 - \cos(q v t)].
\end{equation}
The displacement field in reciprocal space is thus
\begin{equation}
    \Tilde{\mathbf{u}}(\mathbf{q}, t) = -i \Tilde{\phi}_0(q) [1 - \cos(q v t)] \mathbf{q}.
\end{equation}

The functional form of $\phi_0$ is given by the physical mechanism that leads to the distortion. In our model, it is assumed that the dilatation field (i.e., the divergence of the displacement) of the new equilibrium after the excitation is proportional to the concentration of electron-hole pairs~\cite{Gusev1993}. The latter is assumed to follow a spherical Gaussian distribution. Therefore, we may write
\begin{equation}
    \nabla \cdot \mathbf{u}_0(\mathbf{r}) = \nabla^2 \phi_0(\mathbf{r}) = \sigma^{-1} A e^{- r^2/\sigma^2},
    \label{eq:dilatation_field}
\end{equation}
where $A$ is the amplitude of the distortion with units of length and $\sigma$ is the localization size of the electron-hole distribution. With a spherical Fourier transform (see more details in the next section), we can obtain, in reciprocal space,
\begin{equation}
    -q^2 \Tilde{\phi}_0(q) = \frac{\pi^{3/2} A \sigma^2}{V } e^{-\sigma^2 q^2/4},
\end{equation}
where $V$ is a normalization volume which we take to be the volume of the unit cell. Therefore, for $t>0$,
\begin{align}
    \Tilde{\phi}(q, t) =&  -\frac{\pi^{3/2} A \sigma^2 }{V q^2} e^{-\sigma^2 q^2/4} [1 - \cos(q v t)], \label{eq:phi(q,t)_nodamping} \\
    \Tilde{\mathbf{u}}(\mathbf{q}, t) =&  \frac{i\pi^{3/2} A \sigma^2}{V q} e^{-\sigma^2 q^2/4} [1 - \cos(q v t)] \hat{\mathbf{q}}, \label{eq:u(q,t)_nodamping}
\end{align}
where $\hat{\mathbf{q}}$ denotes the unit vector in the direction of $\mathbf{q}$.

If one is interested in the distortion in real space, an inverse spherical Fourier transform can be applied to the results above to obtain:
\begin{align}
     \phi(r, t) = & -\frac{\sqrt{\pi} A \sigma^2}{8  r} \left[ 2 \mathrm{erf} \left( \frac{r}{\sigma} \right) - \mathrm{erf} \left(\frac{r-vt}{\sigma} \right) \right. \nonumber\\
     &\qquad\qquad\qquad \left. - \mathrm{erf} \left( \frac{r+vt}{\sigma} \right)\right] , \label{eq:phi(r,t)_nodamping} \\
     \mathbf{u}(\mathbf{r}, t) = &  \frac{\sqrt{\pi} A \sigma^2 }{8 r^2} \left[ 2 \mathrm{erf} \left( \frac{r}{\sigma} \right) - \mathrm{erf} \left(\frac{r-vt}{\sigma} \right) \right. \nonumber \\
     &\qquad\qquad\qquad\qquad \left. - \mathrm{erf} \left( \frac{r+vt}{\sigma} \right)\right] \hat{\mathbf{r}}  \nonumber \\
     & - \frac{A \sigma}{4 r} \left[ 2 e^{-r^2 / \sigma^2 } - e^{-(r-vt)^2/\sigma^2 } - e^{- (r+vt)^2/\sigma^2} \right] \hat{\mathbf{r}} , \label{eq:u(r,t)_nodamping}
\end{align}
where 
\begin{equation}
    \mathrm{erf}(z) \equiv \frac{2}{\sqrt{\pi}} \int_0^{z} e^{-x^2} dx
\end{equation}
is the error function. The terms containing $(r-vt)/\sigma $ and $(r+vt)/\sigma $ represent outgoing and incoming spherical waves, respectively.

The derivations above have not considered dissipation. In reality, the equilibrium distortion field $s(\mathbf{r})$ decays together with the excited electron cloud, and the phonon modes are damped as well. The time scales of these two processes are not necessarily the same, but in this work, the data is consistent with the two time constants being close to each other. For example, in Fig.~3B in the main text, the experimental data can be described with an overall exponential decay with time. Therefore, we assume that both processes have the same decay time constant, $\tau$. Thus, we may modify the wave equation, Eq.~(\ref{eq:general_wave_equation_reciprocal_space}) into the following form:
\begin{equation}
    - \left( q^2 + \frac{2}{v^2 \tau} \frac{\partial}{\partial t} + \frac{1}{v^2} \frac{\partial^2}{\partial t^2} \right) \Tilde{\phi}(\mathbf{q}, t) = \frac{\pi^{3/2} A \sigma^2}{V } e^{-\sigma^2 q^2/4} e^{-\frac{t}{\tau}} H(t),
\end{equation}
where the term $-(2 v^{-2} \tau^{-1})(\partial\Tilde{\phi}(\mathbf{q}, t)/\partial t) $ accounts for phonon damping, and the term $e^{-t/\tau}$ accounts for the decay of the distortion field. The solution of this equation, with the initial conditions (Eqs.~[\ref{eq:initial_condition_1}, \ref{eq:initial_condition_2}]), is
\begin{align}
    \Tilde{\phi}(q, t) =&  -\frac{\pi^{3/2} A \sigma^2}{V  (q^2-v^{-2}\tau^{-2})} e^{-\sigma^2 q^2/4} \nonumber\\
    &\times \left[1 - \cos\left(q t \sqrt{v^2 - \frac{1}{q^2 \tau^2}} \right) \right]e^{-\frac{t}{\tau}}, \\
    \Tilde{\mathbf{u}}(\mathbf{q}, t) =&  \frac{i\pi^{3/2} A q \sigma^2 }{V (q^2-v^{-2}\tau^{-2})} e^{-\sigma^2 q^2/4} \nonumber\\
    &\times \left[1 - \cos\left(q t \sqrt{v^2 - \frac{1}{q^2 \tau^2}} \right) \right]e^{-\frac{t}{\tau}} \hat{\mathbf{q}}. \\
\end{align}

These results may be simplified under the condition that
\begin{equation}
    q^2 v^2 \tau^2 \gg 1,
\end{equation}
which holds true in our study: for example, with $q=0.004 \times 2\pi\,\si{\per\angstrom}$, $v = \SI{8970}{m/s}$ (for SrTiO$_{3}$), and $\tau \approx \SI{10}{ps}$, we have $q^2 v^2 \tau^2 \approx 500$. Therefore, we may approximate the results above with
\begin{align}
    \Tilde{\phi}(q, t) =&  -\frac{\pi^{3/2} A \sigma^2 }{V q^2} e^{-\sigma^2 q^2/4} \left[1 - \cos\left(q v t \right) \right]e^{-\frac{t}{\tau}}, \label{eq:phi(q,t)} \\
    \Tilde{\mathbf{u}}(\mathbf{q}, t) =&  \frac{i\pi^{3/2} A \sigma^2 }{V q} e^{-\sigma^2 q^2/4} \left[1 - \cos\left(q v t \right) \right]e^{-\frac{t}{\tau}} \hat{\mathbf{q}}, \label{eq:u(q,t)}
\end{align}
which are simply the solution in the undamped case, Eqs.~[\ref{eq:phi(q,t)_nodamping}, \ref{eq:u(q,t)_nodamping}], multiplied by the exponential decay term $e^{-t/\tau}$. Similarly, the solution in real space is given by
\begin{align}
     \phi(r, t) = & -\frac{\pi^{1/2} A \sigma^2}{8  r} e^{-t/\tau} \nonumber\\
     & \quad \times \left[ 2 \mathrm{erf} \left( \frac{r}{\sigma} \right) - \mathrm{erf} \left(\frac{r-vt}{\sigma} \right) - \mathrm{erf} \left( \frac{r+vt}{\sigma} \right)\right], \label{eq:phi(r,t)} \\
     \mathbf{u}(\mathbf{r}, t) = &  \frac{\pi^{1/2} A \sigma^2 }{8 r^2} e^{-t/\tau} \hat{\mathbf{r}} \nonumber\\
     &\quad \times \left[ 2 \mathrm{erf} \left( \frac{r}{\sigma} \right) - \mathrm{erf} \left(\frac{r-vt}{\sigma} \right) - \mathrm{erf} \left( \frac{r+vt}{\sigma} \right)\right] \nonumber \\
     & - \frac{A \sigma}{4 r} e^{-t/\tau} \hat{\mathbf{r}} \nonumber\\
     &\quad \times \left[ 2 e^{-r^2 / \sigma^2 } - e^{-(r-vt)^2/\sigma^2 } - e^{- (r+vt)^2/\sigma^2} \right], \label{eq:u(r,t)}
\end{align}
where $\hat{\mathbf{r}}$ denotes the unit vector in the direction of $\mathbf{r}$.

\subsection{Spherical Fourier transforms}
In this section we show the formulae for Fourier transform pairs in spherical coordinates.

For scalars $\phi(r)$ and $\Tilde{\phi}(q)$:
\begin{align}
    \Tilde{\phi}(q) = & \frac{1}{V} \int_0^\infty r^2 dr \int_0^\pi \sin \theta d\theta \int_0^{2\pi} d\varphi \phi(r) e^{i q r \cos\theta} \nonumber \\ 
            = & \frac{4\pi}{qV} \int_0^\infty \phi(r) r \sin (qr) dr, \\
    \phi(r) = & \frac{V}{2\pi^2 r} \int_0^\infty \Tilde{\phi}(q) q \sin (qr) dq.        
\end{align}
Again, here $V$ is a normalization volume so that $\Tilde{\phi}(q)$ and $\phi(r)$ have the same units. In general, the value of $V$ is arbitrary. For simplicity, in our derivations it is taken to be the unit cell volume.

For vectors $\mathbf{u}(\mathbf{r}) = u(r)\hat{\mathbf{r}}$ and $\Tilde{\mathbf{u}}(\mathbf{q}) = \Tilde{u}(q)\hat{\mathbf{q}}$, note the extra factor of $\cos\theta$ when projecting onto the direction of $\hat{\mathbf{r}}$ or $\hat{\mathbf{q}}$:
\begin{align}
    \Tilde{u}(q) =  & \frac{1}{V} \int_0^\infty r^2 dr \int_0^\pi \sin \theta d\theta \int_0^{2\pi} d\varphi u(r) \cos\theta e^{i q r \cos\theta} \\ 
            = & \frac{4\pi i}{Vq^2} \int_0^\infty u(r) \left[ \sin (qr) - qr \cos(qr) \right] dr, \\
    u(r) = & -\frac{i V}{2\pi^2 r^2} \int_0^\infty \Tilde{u}(q) \left[ \sin (qr) - qr \cos(qr) \right] dq.        
\end{align}

\subsection{Energy in the excited strain field}
The total energy deposited to the LA phonon fields can be calculated from the momentum-resolved LA displacements by integrating over all modes.  The energy per mode is 

\begin{equation}
   \Tilde{W}(q) =  \frac{1}{2} cN m \omega^2(q)|\tilde{u}(q)|^2 = \frac{1}{2} cN m v^2q^2 |\tilde{u}(q)|^2,\label{eq:Energy_density_derivation_1}
\end{equation}
where $v$ is the speed of sound, $q$ is the magnitude of the wavevector, $m$ is the total mass of atoms in the unit cell, and
\begin{equation}
    |\tilde{u}(q)| = \frac{\pi^{3/2} A \sigma^2 }{V q} e^{-\sigma^2 q^2/4}
\end{equation}
is the maximum mode displacement at a given wavevector; see Eq.~(\ref{eq:u(q,t)}). Since we take into account an exponential decay in time, $\Tilde{W}(q)$ thus represents the phonon energy at $t=0$. Integrating over all wavevectors and dividing by the total volume $NV$, we obtain the energy density:
\begin{align}
    U_p & = \frac{1}{NV} \frac{V}{(2\pi)^3} \int_0^{\infty} 4\pi q^2 dq \Tilde{W}(q) \\
    & = \frac{1}{8 \pi^3 N} \int_0^{\infty} 4\pi q^2 dq \cdot \frac{1}{2} cN m v^2 q^2 \left( \frac{\pi^{3/2} A \sigma^2 }{V q} \right)^2 e^{-\sigma^2 q^2/2} \\
    & = \frac{\pi c m v^2 A^2 \sigma^4}{4 V^2} \int_0^{\infty} q^2 e^{-\sigma^2 q^2/2} dq \\
    & = \frac{\pi^{3/2} c m v^2 A^2 \sigma}{4 \sqrt{2} V^2 }. \label{eq:Up_result}
\end{align}



The same result can be obtained via calculations in real space. Because there is no shear, the elastic energy density of the distortion field is~\cite{Chandrasekharaiah1994}:
\begin{align}
    W &= \frac{1}{2} (\lambda_L+2\mu_L) (\varepsilon_{11}+\varepsilon_{22}+\varepsilon_{33})^2 \nonumber \\
    &\quad - 2\mu_L (\varepsilon_{11}\varepsilon_{22} + \varepsilon_{22}\varepsilon_{33} + \varepsilon_{33}\varepsilon_{11}) \\
    &= \frac{1}{2} (\lambda_L+2\mu_L)(\nabla \cdot \mathbf{u})^2   - 2\mu_L (\varepsilon_{11}\varepsilon_{22} + \varepsilon_{22}\varepsilon_{33} + \varepsilon_{33}\varepsilon_{11}),
\end{align}
where $\varepsilon_{11,22,33}$ are the diagonal elements of the strain tensor; $\lambda_L$ and $\mu_L$ are the material's Lam\'e parameters, which are related to the longitudinal sound speed by
\begin{equation}
    \lambda_L + 2\mu_L = \frac{m}{V} v^2.
\end{equation}

The amplitude profile in $q$ corresponds to the amplitude of the time-independent term in Eq.~(\ref{eq:u(r,t)}):
\begin{equation}
    \mathbf{u}_0(\mathbf{r}) = \left[ \frac{\pi^{1/2} A \sigma^2 }{4 r^2} \mathrm{erf} (r/\sigma) - \frac{A \sigma}{2r}e^{-r^2/\sigma^2}\right] \hat{\mathbf{r}}. 
\end{equation}
Thus we can easily obtain the strain tensor due to a single defect:
\begin{align}
    \varepsilon_{11} &= \varepsilon_{rr} = \frac{d u_0}{dr}, \\
    \varepsilon_{22} &= \varepsilon_{\theta\theta} = \frac{u_0}{r}, \\
    \varepsilon_{33} &= \varepsilon_{\phi\phi} =  \frac{u_0}{r}.
\end{align}
Therefore, the energy density of the excitations in the system is
\begin{align}
    U_p &= \frac{cN}{NV} \int_0^\infty 4\pi r^2 W dr  \nonumber \\
    &= \frac{4\pi c}{V} \int_0^\infty \left[ \frac{1}{2} (\lambda_L+2\mu_L) r^2 (\nabla \cdot \mathbf{u}_0)^2 \nonumber \right.  \nonumber \\
    &\qquad\qquad\qquad  - \left. 2\mu_L \left( 2 r u_0 \frac{du_0}{dr} + u_0^2 \right) \right] dr \nonumber \\
    &= \frac{4\pi c}{V} \left[\int_0^\infty \frac{m v^2}{2 V \sigma^2} r^2 A^2 e^{-2r^2 / \sigma^2} dr - \left. (2\mu_L u_0^2 r) \right|_{r=0}^\infty \right] \\
    &= \frac{\pi^{3/2} c m v^2 A^2 \sigma}{4 \sqrt{2} V^2 }.
    \label{eq:Up_expression}
\end{align}

\subsection{\label{sec:obtaining_conversion_efficiency}Obtaining the energy conversion efficiency}
In this section, we present detailed derivations of how the conversion efficiency (from deposited X-ray energies to phonon energies) can be obtained by fitting the data with the model described above. The only additional assumption is that the term $cA^2$, which describes the concentration and amplitude of the excitations, is proportional to the fluence of the pump pulse at any point in the sample. As will be shown, this is expected given the bi-linearity of the pump-probe signal demonstrated in the main text.

\begin{figure}
    \centering
    \includegraphics[width=0.95\columnwidth]{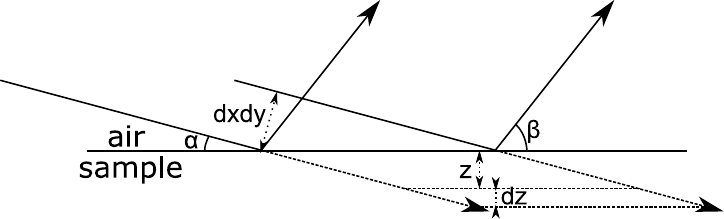}
    \caption{Schematic diagram showing the geometric parameters used in the derivations. Note that the coordinates $x,y$ are transverse to the beam axis, while the coordinate $z$ is in the direction normal to the sample surface.}
    \label{fig:geometric_parameters}
\end{figure}

We begin by considering the scattering from a single pair of pump-probe pulses. Assuming that, at the sample position, the probe and pump beams have transverse fluence profiles $\Phi_1(x,y)$ and $\Phi_2(x,y)$. Let $\mu$ denote the X-ray linear attenuation coefficient. Since we work in grazing geometry, let $\alpha$ denote the grazing angle, and $\beta$ the angle of the outgoing wave (see Fig.~\ref{fig:geometric_parameters}). As in the main text, we use $\mathbf{Q}$ to denote the scattering wavevector, $\mathbf{G}$ the reciprocal lattice vector, and $\mathbf{q} \equiv \mathbf{Q}-\mathbf{G}$ the deviation from the Bragg peak which corresponds to the phonon wave vector. Then, the thermal diffuse scattering to the first order can be written as~\citep{Xu2005a}:
\begin{align}
    I_0(\mathbf{Q}) =& \frac{\hbar}{2} \sum_{i} \frac{1}{\omega_{\mathbf{q},i}} \coth \left( \frac{\hbar \omega_{\mathbf{q},i}}{2 k_B T} \right) \left|\sum_j F_j(\mathbf{G}) \left( \mathbf{Q}\cdot\frac{\mathbf{\epsilon}_{i,\mathbf{q},j}}{{\sqrt{m_j}}} \right) \right|^2 \nonumber \\
    &\qquad \times \int_{-\infty}^\infty \int_{-\infty}^\infty \frac{dxdy}{\sin\alpha} \int_0^\infty dz \exp\left(- \frac{\mu z}{\sin\beta} \right) I_e n,
\label{eq:TDS_1st_order}
\end{align}
where $n$ is the number density of the unit cell, $k_B$ is the Boltzmann constant, and $T=\SI{300}{K}$ is the sample temperature. The $\exp (- \mu z/\sin\beta )$ term accounts for the attenuation of the outgoing beam. The sum $\sum_j$ is over all atoms in a unit cell; $m_j$ is the mass of atom $j$, and the structure factor of atom $j$ is defined as
\begin{equation}
F_j(\mathbf{G}) \equiv f_j e^{-M_j} e^{-i \mathbf{G} \cdot \mathbf{\tau}_j},
\end{equation}
where $f_j$ is the form factor, $e^{-M_j}$ the Debye-Waller factor, and $\mathbf{\tau}_j$ the position of the atom in the unit cell.
The sum $\sum_i$ is over all phonon modes; $\omega_{\mathbf{q},i}$ is the angular frequency and $\mathbf{\epsilon}_{i,\mathbf{q},j}$ the eigenvector of phonon mode $i$. $z$ is the penetration depth into the sample (see Fig.~\ref{fig:geometric_parameters}). $I_e$ is the scattering from a single electron; it can be re-written as 
\begin{equation}
    I_e=\Phi_\text{inc}(x,y,z) \mathcal{S}_e,
\end{equation}
where $\Phi_\text{inc}(x,y,z)$ is the total incident fluence at coordinate $(x,y,z)$, and $\mathcal{S}_e$ is a scattering strength taking into account X-ray polarization and detector solid angle; see Ref.~\citep{Xu2005a}. Taking into account the spatial profile of the X-ray beams as well as their attenuation in the sample giving rise to a factor of $\exp(- \mu z/\sin\alpha )$, the equation above becomes:
\begin{widetext}
\begin{equation} \label{eq:I0_full_1}
    \begin{split}
    I_0(\mathbf{Q}) =&  \frac{\hbar n \mathcal{S}_e}{2} \sum_{i} \frac{1}{\omega_{\mathbf{q},i}} \coth \left( \frac{\hbar \omega_{\mathbf{q},i}}{2 k_B T} \right) \left| \sum_j F_j(\mathbf{G}) \left(\mathbf{Q}\cdot\frac{\mathbf{\epsilon}_{i,\mathbf{q},j}}{{\sqrt{m_j}}} \right) \right|^2 \\
    & \quad \times \int_{-\infty}^\infty \int_{-\infty}^\infty \frac{dxdy}{\sin\alpha } \int_0^\infty dz [\Phi_1(x,y)+\Phi_2(x,y)] \exp\left(-\frac{\mu z}{\sin\alpha} - \frac{\mu z}{\sin\beta} \right).
    \end{split}
\end{equation}
\end{widetext}
Noting that the integral of the fluence is the pulse energy,
\begin{equation}
    \int_{-\infty}^\infty \int_{-\infty}^\infty dxdy \Phi_{1,2}(x,y) = \mathcal{E}_{1,2},
\end{equation}
we may re-write Eq.~(\ref{eq:I0_full_1}) as 
\begin{align} \label{eq:I0_full_2}
    I_0(\mathbf{Q}) =& \frac{\hbar n \mathcal{S}_e}{2 \mu} (\mathcal{E}_1 + \mathcal{E}_2) \sum_{i} \frac{1}{\omega_{\mathbf{q},i}} \coth \left( \frac{\hbar \omega_{\mathbf{q},i}}{2 k_B T} \right) \nonumber \\
    & \quad \times \left| \sum_j F_j(\mathbf{G}) \left( \mathbf{Q}\cdot\frac{\mathbf{\epsilon}_{i,\mathbf{q},j}}{{\sqrt{m_j}}} \right) \right|^2 
    \left( 1 + \frac{\sin\alpha}{\sin\beta} \right)^{-1} .
\end{align}
As expected, $I_0(\mathbf{Q})$ is proportional to the summed pulse energy $\mathcal{E}_1+\mathcal{E}_2$. As in Eq.~(\ref{eq:I=I0+DeltaI}), we may write $I_0(\mathbf{Q}) = S_0(\mathbf{Q}) (\mathcal{E}_1 + \mathcal{E}_2)$, where $S_0(\mathbf{Q})$ is independent of $\mathcal{E}_{1,2}$.

The change in diffuse scattering intensity due to the distortions can be derived in a similar way as for the Huang diffuse scattering due to static defects~\cite{dederichs_theory_1973}. The results can be written as:
\begin{align}
    \Delta I(\mathbf{Q},t) \approx & \int_{-\infty}^\infty \int_{-\infty}^\infty  n \frac{dxdy}{\sin\alpha } \int_0^\infty dz c I_e \exp\left(- \frac{\mu z}{\sin\beta} \right) \nonumber\\
    & \quad \times \left| \sum_j F_j(\mathbf{G}) \right|^2 \left| \mathbf{G} \cdot \Tilde{\mathbf{u}}(\mathbf{q},t) \right|^2 \\
    = &  \int_{-\infty}^\infty \int_{-\infty}^\infty n \frac{dxdy}{\sin\alpha } \int_0^\infty dz c \mathcal{S}_e \Phi_1(x, y) \nonumber \\
    & \quad \times \exp\left(-\frac{\mu z}{\sin\alpha} - \frac{\mu z}{\sin\beta} \right) \left| \sum_j F_j(\mathbf{G}) (\mathbf{G} \cdot \hat{\mathbf{q}}) \right|^2 \nonumber \\
    & \quad \times \frac{\pi^3 A^2 \sigma^4 }{V^2 q^2} e^{-\frac{\sigma^2 q^2}{2}} \left[1 - \cos\left(q v t \right) \right]^2 e^{-\frac{2 t}{\tau}}, 
\end{align} 
where in the second step we have used the results from the model, Eq.~(\ref{eq:u(q,t)}). Note that here the incident fluence $\Phi_\text{inc}$ includes only the probe beam, $\Phi_1$. As mentioned above, the effect of the pump pulse on the sample is reflected in the term $cA^2$, which varies with the spatial coordinates $(x,y,z)$ and is assumed to be proportional to the pump fluence:
\begin{equation} \label{eq:kappa_definition}
    cA^2 = \kappa \Phi_\text{pump} = \kappa \Phi_2(x,y) \exp\left(- \frac{\mu z}{\sin\alpha} \right) ,
\end{equation}
where $\kappa$ is a conversion coefficient. Thus,
\begin{widetext}
\begin{equation} \label{eq:DeltaI_full_1}
\begin{split}
    \Delta I(\mathbf{Q},t) = & \frac{\pi^3 \kappa \sigma^4 n \mathcal{S}_e}{V^2 q^2} e^{-\frac{\sigma^2 q^2}{2}} \left[1 - \cos\left(q v t \right) \right]^2 e^{-\frac{2 t}{\tau}}\\
    & \times \int_{-\infty}^\infty \int_{-\infty}^\infty  \frac{dxdy}{\sin\alpha } \int_0^\infty dz \Phi_1(x,y) \Phi_2(x,y) \exp\left(-\frac{2 \mu z}{\sin\alpha} - \frac{\mu z}{\sin\beta} \right) \left| \sum_j F_j(\mathbf{G}) (\mathbf{G} \cdot \hat{\mathbf{q}}) \right|^2. 
\end{split}
\end{equation}
    
\end{widetext}
As will be discussed in the section ``Overlap correction'' below, the beam profiles may change during a delay scan due to motor movements. However, at a given delay, we may assume that the spatial profiles of the beams remain the same for all shots. In other words, we may write $\Phi_{1,2}(x,y) = \mathcal{E}_{1,2} \phi_{1,2}(x,y)$, where $\phi_{1,2}(x,y)$ do not vary between shots and $\iint \phi_{1,2}(x,y) dx dy = 1$. Then, we define the overlap factor:
\begin{equation} \label{eq:overlap_factor_definition}
    \mathcal{O}(t) \equiv 4\pi \sigma_b^2 \int_{-\infty}^\infty \int_{-\infty}^\infty \phi_1(x,y) \phi_2(x,y) dxdy.
\end{equation}
The prefactor $4\pi \sigma_b^2$ represents the area of the beam and makes $\mathcal{O}(t)$ a unitless quantity. $\sigma_b$ represents the size of the beam and, in case of a Gaussian beam, it is taken to be the standard deviation of the Gaussian (see the section ``Overlap correction'' below). Now, we can rewrite Eq.~(\ref{eq:DeltaI_full_1}) as:
\begin{equation} \label{eq:DeltaI_full_2}
\begin{split}
    \Delta I(\mathbf{Q},t) =& \frac{\pi^2 \kappa \sigma^4 n \mathcal{S}_e}{4 \sigma_b^2 V^2 q^2 \mu} e^{-\sigma^2 q^2/2} \left[1 - \cos\left(q v t \right) \right]^2 e^{-2 t/\tau} \mathcal{E}_1 \mathcal{E}_2 \\
    & \quad \times \left| \sum_j F_j(\mathbf{G}) (\mathbf{G} \cdot \hat{\mathbf{q}}) \right|^2 \mathcal{O}(t)
    \left( 2+ \frac{\sin\alpha}{\sin\beta} \right)^{-1} .     
\end{split}
\end{equation}
As expected, this pump-probe signal is bi-linear in the pump and probe pulse energies. As in the main text, we may write $\Delta I(\mathbf{Q}, t) = C(\mathbf{Q}, t) \mathcal{O}(t) \mathcal{E}_1 \mathcal{E}_2$, where $C(\mathbf{Q}, t)$ is independent of $\mathcal{E}_{1,2}$. We have also isolated the overlap correction factor $\mathcal{O}(t)$, a purely geometrical effect due to experimental conditions, from the physically relevant quantity $C(\mathbf{Q}, t)$.

Experimentally, we measure the total intensity $I(\mathbf{Q}, t) = I_0(\mathbf{Q}) + \Delta I (\mathbf{Q}, t)$ together with the pulse energies $\mathcal{E}_1$, $\mathcal{E}_2$ for each shot. Let $s$ be the index of a shot, then the summed intensity is
\begin{equation}
\begin{split}
    \sum_s^{\text{all shots}} I(\mathbf{Q}, t) =& S_0(\mathbf{Q}) \sum_s^{\text{all shots}} (\mathcal{E}_1^{(s)} + \mathcal{E}_2^{(s)}) \\
    &+ C(\mathbf{Q}, t) \mathcal{O}(t) \sum_s^{\text{all shots}} \mathcal{E}_1^{(s)} \mathcal{E}_2^{(s)}.    
\end{split}
\end{equation}
Then, we normalize it by the summed pulse energies:
\begin{align}
    I^{\text{norm}} (\mathbf{Q}, t) \equiv & \left. \sum_s I(\mathbf{Q}, t) \middle/ \sum_s (\mathcal{E}_1^{(s)} + \mathcal{E}_2^{(s)}) \right. \nonumber \\
    =& S_0(\mathbf{Q}) + C(\mathbf{Q}, t) \mathcal{O}(t) \frac{\sum_s \mathcal{E}_1^{(s)}\mathcal{E}_2^{(s)}}{\sum_s (\mathcal{E}_1^{(s)} + \mathcal{E}_2^{(s)})}.
\end{align}
As shown in the main text, there is no pump-probe signal at $t=0$, so the term $S_0(\mathbf{Q})$ may be replaced by $I^{\text{norm}} (\mathbf{Q}, t=0)$. Hence,
\begin{widetext}
\begin{align}
    & \left[ \frac{I^{\text{norm}} (\mathbf{Q}, t)}{I^{\text{norm}} (\mathbf{Q}, t=0)} - 1 \right] \left[ \sum_s (\mathcal{E}_1^{(s)} + \mathcal{E}_2^{(s)}) \middle/ \sum_s \mathcal{E}_1^{(s)}\mathcal{E}_2^{(s)} \right] [\mathcal{O}(t)]^{-1} \\
    = & C(\mathbf{Q}, t) / S_0(\mathbf{Q}) \\
    = & \left. \frac{\pi^2 \kappa \sigma^4 n \mathcal{S}_e}{4\sigma_b^2 V^2 q^2 \mu} e^{-\frac{\sigma^2 q^2}{2}} \left[1 - \cos\left(q v t \right) \right]^2 e^{-\frac{2 t}{\tau}} \left| \sum_j F_j(\mathbf{G}) (\mathbf{G} \cdot \hat{\mathbf{q}}) \right|^2 
    \left( 2+ \frac{\sin\alpha}{\sin\beta} \right)^{-1} \right/ \nonumber \\
    & \frac{\hbar n \mathcal{S}_e}{2 \mu} \sum_{i} \frac{1}{\omega_{\mathbf{q},i}} \coth \left( \frac{\hbar \omega_{\mathbf{q},i}}{2 k_B T} \right) \left| \sum_j F_j(\mathbf{G}) \left(\mathbf{Q}\cdot\frac{\mathbf{\epsilon}_{i,\mathbf{q},j}}{{\sqrt{m_j}}} \right) \right|^2 
    \left( 1 + \frac{\sin\alpha}{\sin\beta} \right)^{-1} \\
    = & \frac{\pi^2 \kappa \sigma^4}{2\hbar \sigma_b^2 V^2 q^2} e^{-\frac{\sigma^2 q^2}{2}} \left[1 - \cos\left(q v t \right) \right]^2 e^{-\frac{2 t}{\tau}} \frac{\left| \sum_j F_j(\mathbf{G}) (\mathbf{G} \cdot \hat{\mathbf{q}}) \right|^2}{\sum_{i} \omega_{\mathbf{q},i}^{-1} \coth \left( \frac{\hbar \omega_{\mathbf{q},i}}{2 k_B T} \right) \left| \sum_j F_j(\mathbf{G}) \left(\mathbf{Q}\cdot\frac{\mathbf{\epsilon}_{i,\mathbf{q},j}}{{\sqrt{m_j}}} \right) \right|^2 } \left(\frac{ 1 + \frac{\sin\alpha}{\sin\beta} }{2+ \frac{\sin\alpha}{\sin\beta} }\right)
    \\
    \approx & \frac{\pi^2 \kappa \sigma^4}{4 \hbar \sigma_b^2 V^2 q^2} e^{-\frac{\sigma^2 q^2}{2}} \left[1 - \cos\left(q v t \right) \right]^2 e^{-\frac{2 t}{\tau}} \frac{\left| \sum_j F_j(\mathbf{G}) (\mathbf{G} \cdot \hat{\mathbf{q}}) \right|^2}{\sum_{i} \omega_{\mathbf{q},i}^{-1} \coth \left( \frac{\hbar \omega_{\mathbf{q},i}}{2 k_B T} \right) \left| \sum_j F_j(\mathbf{G}) \left(\mathbf{Q}\cdot\frac{\mathbf{\epsilon}_{i,\mathbf{q},j}}{{\sqrt{m_j}}} \right) \right|^2 }
    \label{eq:pump_probe_fit_function_1}
\end{align}
\end{widetext}
where in the last step we have used approximations given that $\sin\alpha / \sin\beta \ll 1$. 

The physical quantity of interest is the ratio between the deposited energy density, $U_d$, and the phonon energy density, $U_p$. The former is simply $U_d = \mu_\text{pe} \Phi_\text{pump}$, where $\mu_\text{pe}$ is the x-ray photoelectric absorption coefficient~\cite{xdb}. Thus, combining Eqs.~[\ref{eq:kappa_definition}, \ref{eq:Up_result}, \ref{eq:pump_probe_fit_function_1}], we obtain
\begin{widetext}
\begin{align}
    & \left[ \frac{I^{\text{norm}} (\mathbf{Q}, t)}{I^{\text{norm}} (\mathbf{Q}, t=0)} - 1 \right] \left[ \sum_s (\mathcal{E}_1^{(s)} + \mathcal{E}_2^{(s)}) \middle/ \sum_s \mathcal{E}_1^{(s)}\mathcal{E}_2^{(s)} \right] [\mathcal{O}(t)]^{-1}
    \label{eq:pump_probe_signal_experiment} \\
    = & \frac{(2\pi)^{1/2} \mu_\text{pe} \sigma^3 }{\hbar  \sigma_b^2  m v^2 q^2} \left( \frac{U_p}{U_d} \right) e^{-\frac{\sigma^2 q^2}{2}} \left[1 - \cos\left(q v t \right) \right]^2 e^{-\frac{2 t}{\tau}} \frac{\left| \sum_j F_j(\mathbf{G}) (\mathbf{G} \cdot \hat{\mathbf{q}}) \right|^2}{\sum_{i} \omega_{\mathbf{q},i}^{-1} \coth \left( \frac{\hbar \omega_{\mathbf{q},i}}{2 k_B T} \right) \left| \sum_j F_j(\mathbf{G}) \left(\mathbf{Q}\cdot\frac{\mathbf{\epsilon}_{i,\mathbf{q},j}}{{\sqrt{m_j}}} \right) \right|^2 }. \label{eq:pump_probe_fit_function_2}
\end{align}
\end{widetext}
Therefore, by calculating the pump-probe signal in Eq.~(\ref{eq:pump_probe_signal_experiment}) from experimental data, and fitting it with Eq.~(\ref{eq:pump_probe_fit_function_2}), we can extract the localization size $\sigma$ and the conversion coefficient $U_p/U_d$. Specifically, we may define a pre-factor $\mathcal{F}$:
\begin{equation}
    \mathcal{F} \equiv \frac{(2\pi)^\frac{1}{2} \mu_\text{pe} }{\hbar  \sigma_b^2  m v^2 q^2} \frac{\left| \sum_j F_j(\mathbf{G}) \right|^2}{\sum_{i} \omega_{\mathbf{q},i}^{-1} \coth \left( \frac{\hbar \omega_{\mathbf{q},i}}{2 k_B T} \right) \left| \sum_j F_j(\mathbf{G}) \left(\mathbf{Q}\cdot\frac{\mathbf{\epsilon}_{i,\mathbf{q},j}}{{\sqrt{m_j}}} \right) \right|^2 },
    \label{eq:pre_factor_F}
\end{equation}
which includes: geometric factors (e.g., the beam size and the pump-probe overlap factor), known constants (e.g., the x-ray linear absorption coefficient), DFT results (e.g., phonon mode frequencies), all of which are independent from parameters of the model. Thus, the pre-factor $\mathcal{F}$ can be calculated for any given $\mathbf{q}, t$ and is fixed during the data fitting. We can then re-write the equation above as
\begin{align}
    &\left[ \frac{I^{\text{norm}} (\mathbf{Q}, t)}{I^{\text{norm}} (\mathbf{Q}, t=0)} - 1 \right] \left[ \frac{\sum_s (\mathcal{E}_1^{(s)} + \mathcal{E}_2^{(s)})}{\sum_s \mathcal{E}_1^{(s)}\mathcal{E}_2^{(s)}} \right] [\mathcal{O}(t)]^{-1} \nonumber \\
    =& \frac{C(\mathbf{Q}, t)}{S_0(\mathbf{Q})} \nonumber \\
    =& \mathcal{F} \sigma^3 \left( \frac{U_p}{U_d} \right) e^{-\frac{\sigma^2 q^2}{2}} \left[1 - \cos\left(q v t \right) \right]^2 e^{-\frac{2 t}{\tau}} \left|\mathbf{G} \cdot \hat{\mathbf{q}} \right|^2,
\end{align}    
and fit the data by tuning the parameters $\tau$, $\sigma$, and $U_p/U_d$.

\section{Additional experimental methods}
\subsection{\label{sec:overlap_correction}Overlap correction}
The time delay between the two pulses is adjusted via two symmetric linear motions in the delay branch which change the distance between the inner crystals (i.e., C$_1$ and C$_4$ in Fig.~1A in the main text) and the outer crystals (C$_2$ and C$_3$). In order to perform a continuous scan of the delay, the straightness of the linear stages needs to meet two requirements:
\begin{enumerate}
    \item The orientation errors of the outer crystals caused by this linear motion should be well below the Darwin width of the Bragg reflection, \SI{17}{\micro\radian} in this case, to maintain the photon throughput.
    \item The angular errors of the exit beam from the delay branch should be sufficiently small so that the two output beams remain focused and overlapped at the sample location. Note that in this experiment, with a focal size of \SI{20}{\micro\meter}  and a focal length of \SI{3.3}{m}, angular errors on the order of \SI{6}{\micro\radian} would lead to the complete loss of overlap between the two beams.
\end{enumerate}
Although the planar air-bearing-based mechanism used for the linear motion~\cite{Zhu2017_Development} meets the first requirement, it is still difficult to achieve a sub-\si{\micro\radian} level straightness required by the second one. On the other hand, the angular errors of these air-bearing-based linear motions are repeatable on the sub-\si{\micro\radian} level. Therefore, the following calibration routine has been implemented to partially correct for the angular errors:
\begin{enumerate}
    \item First, we measure changes in the horizontal and vertical position of the beam due to the angular movements during the delay scan. This is done using the high-resolution beam profile monitor at the sample location. 
    \item Then, we calibrate the relation between the $\theta$ and $\chi$ motion of crystal C$_4$ and the horizontal and vertical movement of the beam at the sample location using the profile monitor. 
    \item Next, we build a lookup table for $\theta$ and $\chi$ values to compensate for the angular motion measured in step 1.
    \item Finally, we perform the delay scan and at each time point, using the values of $\theta$ and $\chi$ in the lookup table.  
\end{enumerate}

\begin{figure}
\centering
\includegraphics{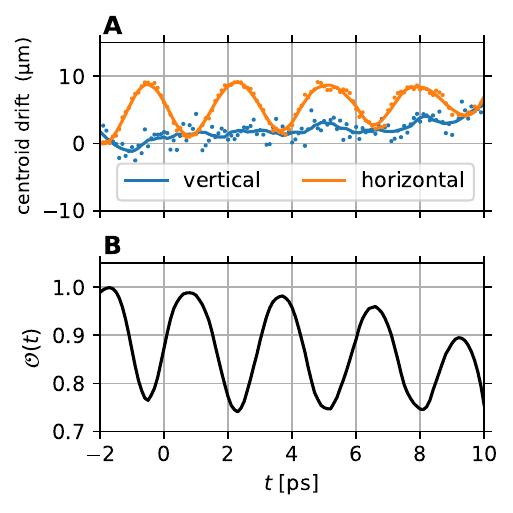}
\caption{(A) Measured drifts in the beam center position as a function of delay $t$ between the two X-ray pulses after the overlap correction. The dots are data points at each delay, which are then smoothed using a Savitzky–Golay filter and the results are shown as the curves. (B) The overlap correction factor, $\mathcal{O}(t)$, calculated with Eq.~(\ref{eq:overlap_factor_calculation}) using the values in (A).}
\label{fig:overlap_correction}
\end{figure}

Shown in Fig.~\ref{fig:overlap_correction}A is the movement of the focused beam from the delay branch measured at the sample position using the beam profile monitor while the delay is scanned from \SIrange{-2}{10}{ps} after the angular error correction. Limited by the resolution of the $\theta$ and $\chi$ motions, we can only correct the angular errors to some extent. During the experiment, the overlap is optimized at the most negative delay, $t=\SI{-2}{ps}$, so the change in the centroid position is calculated with respect to its position at \SI{-2}{ps}.

To calculate the overlap correction factor, $\mathcal{O}(t)$, defined in Eq.~(\ref{eq:overlap_factor_definition}), we assume that the two beams are both Gaussian in shape. Since the full-width at half-maximum (FWHM) of the beams are measured to be \SI{20}{\micro\meter}, the standard deviation of the Gaussian is thus $\sigma_b = \SI{8.49}{\micro\meter}$. Let $D$ denote the distance between the centroids of the two beams. We may choose the coordinate system so that the centroids are located at $(x,y) = (\pm D/2, 0)$. Hence, the overlap factor is
\begin{align}
    \mathcal{O}(t) =& 4\pi\sigma_b^2 \int_{-\infty}^\infty \int_{-\infty}^\infty dxdy \frac{1}{2\pi \sigma_b^2} \exp \left[ -\frac{(x+\frac{D}{2})^2 + y^2}{2\sigma_b^2} \right] \nonumber \\
    &\qquad\qquad  \times \frac{1}{2\pi \sigma_b^2}  \exp \left[ -\frac{(x-\frac{D}{2})^2 + y^2}{2\sigma_b^2} \right] \\
    =& \exp \left( -\frac{D^2}{4\sigma_b^2} \right) \label{eq:overlap_factor_calculation}
\end{align}
This factor is calculated and plotted in Fig.~\ref{fig:overlap_correction}B and is taken into account for further analyses on the time-resolved signal.

\subsection{\label{sec:diode_calibration}Diode calibration}
\begin{figure}
    \centering
    \includegraphics{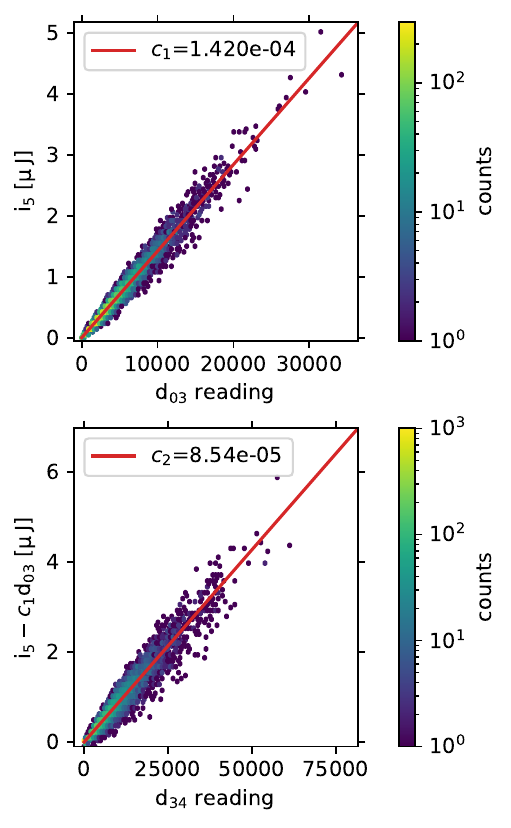}
    \caption{Diode calibration. Top panel: we obtain the coefficient $c_1$ which converts d$_{03}$ reading (variable-delay branch) to pulse energy onto the sample by blocking the fixed-delay branch and fitting the correlation between d$_{03}$ and i$_5$ readings. Bottom panel: we obtain the coefficient $c_2$ which converts d$_{34}$ reading (fixed-delay branch) to pulse energy onto the sample by leaving both branches open and fitting the correlation between d$_{34}$ and $\mathrm{i}_5 - c_1 \mathrm{d}_{03}$.}
    \label{fig:diode_calibration}
\end{figure}

Figure~\ref{fig:diode_calibration} shows the calibration for the diode readings. The intensity monitor i$_5$, which is placed before the sample, is calibrated separately and the readings are in units of \micro J. We use the reading to calibrate the diodes d$_{03}$ and d$_{34}$ in the following way: In one scan, we block the fixed-delay branch and obtain the coefficient $c_1$ that converts d$_{03}$ reading (i.e., intensity from the variable-delay branch) into i$_5$ reading, as shown in the top panel of Fig.~\ref{fig:diode_calibration}. In another scan, we leave both branches open and, knowning the coefficient $c_1$, obtain the coefficient $c_2$ that converts d$_{34}$ reading (i.e., intensity from the fixed-delay branch) into i$_5$ reading, as shown in the bottom panel of Fig.~\ref{fig:diode_calibration}. In this way, we can obtain the energy delivered onto the sample in units of \micro J from each branch, $\mathcal{E}_1=c_1\mathrm{d}_{03}$ and $\mathcal{E}_2=c_2\mathrm{d}_{34}$.

\section{\label{sec:additional_data}Additional data}
\begin{figure*}
\centering
\includegraphics
{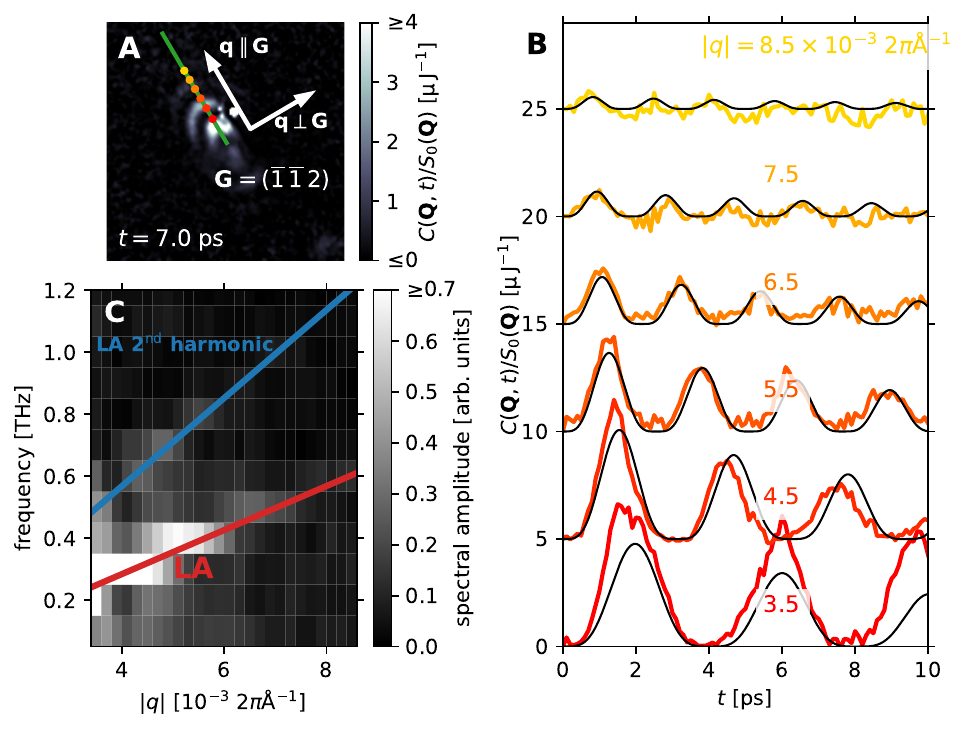}
\caption{Measured x-ray pump x-ray probe signal $C(\mathbf{Q},t)/S_0(\mathbf{Q})$ in KTaO$_{3}$.  
    (A) $C(\mathbf{Q},t)/S_0(\mathbf{Q})$ at $t=7.0$ ps. The green line shows the direction $\mathbf{q}\parallel\mathbf{G}$, which coincides with the direction of the largest intensity modulation. (B) The time dependence of $C(\mathbf{Q},t)/S_0(\mathbf{Q})$ at selected wavevectors $\mathbf{q}$ along the red line in (A). The corresponding locations on the detector are indicated as colored dots in (A). An offset is added between traces of different $|q|$ values for clarity; $C(\mathbf{Q},t)/S_0(\mathbf{Q})$ is zero at $t=0$. The black lines are fit results. (C) Fourier transform spectral amplitudes along the direction of the red line in (A). The red and blue lines show the dispersion of the LA phonon and the LA second harmonic obtained from DFT calculations.
}\label{fig:KTO1}
\end{figure*}

\begin{figure*}
\centering
\includegraphics
{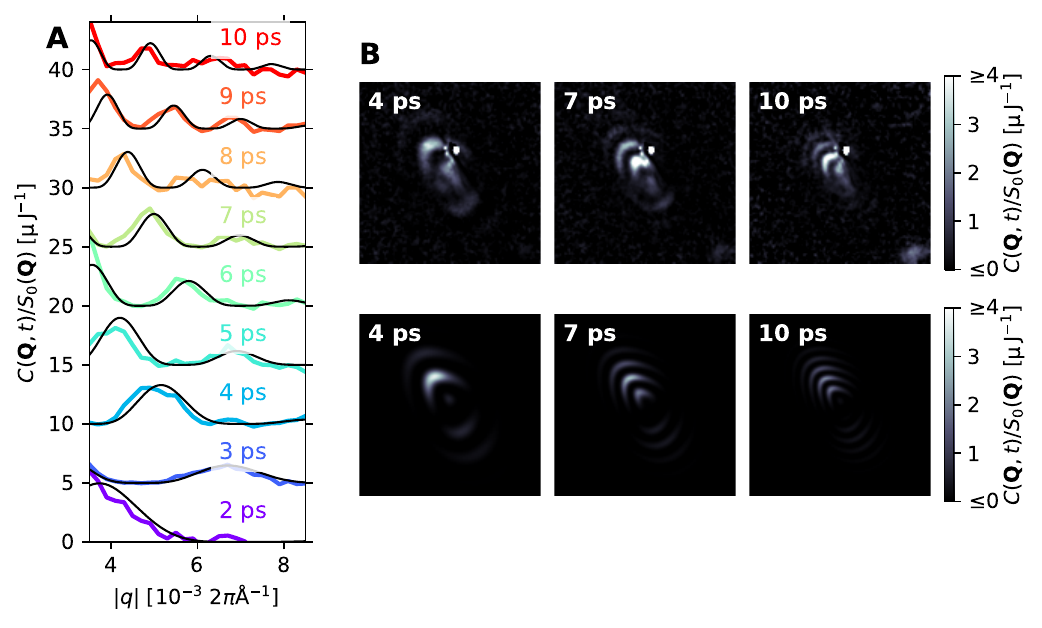}
\caption{Fit results for KTaO$_{3}$.  
    (A)  The pump-probe signal $C(\mathbf{Q},t)/S_0(\mathbf{Q})$ along the $\mathbf{q}$ linecut in Fig.~\ref{fig:KTO1}A, at selected delay times. Black lines are the simulations, colored lines are the data.
    (B) $C(\mathbf{Q},t)/S_0(\mathbf{Q})$ on the detector image (at $t=4,7,10$\,ps) are compared between the experiment (first row) and the simulations (second row).
} \label{fig:KTO2}
\end{figure*}

Figures~\ref{fig:KTO1} and \ref{fig:KTO2} present data on KTaO$_3$, in the same format as Figs.~\ref{fig:time_traces} and \ref{fig:fit_results} in the main text for SrTiO$_3$.



\section{\label{sec:energy_conversion}Estimated amplitude of strain waves}
If we assume that the coupling is only through deformation potential and that one photon creates one spherical wave, then each photon will lead to a uniform electron band shift of the amount $\Delta E$ in the excited volume $\Omega=NV=\frac{4\pi}{3}\sigma_e^3$. $V$ is the unit cell volume and $N $ is the number of unit cells excited. $\sigma_e$ is the radius of the electron cloud, which is allowed to be different from $\sigma$. The photon of energy $h\nu$ injects a number of carriers $\Delta N=h\nu/E_{\text{gap}}$ and leads to a change in the chemical potential $\Delta E=\Delta N E_{\text{gap}}/N$ where $N$ the total number of unit cells in the excited volume $\Omega$ or the number of electrons within one single band is defined through $\Omega=NV=\frac{4\pi}{3}\sigma_e^3$. We then have $\Delta E=\frac{h\nu}{N}$.  
The induced strain is determined by $\Delta E=\Xi\varepsilon$ where  $\Xi$ is the deformation potential, Therefore, the uniform strain $\varepsilon$, which is incurred by the incident photon, satisfies the relation $\frac{\varepsilon}{h\nu}=\frac{V}{\Xi\Omega}=\frac{3V}{4\pi \Xi\sigma_e^3}$. 
The signal $\frac{[C(\mathbf{Q},t)/S_0(\mathbf{Q})]_{\text{STO}}}{[C(\mathbf{Q},t)/S_0(\mathbf{Q})]_{\text{GaAs}}}\approx \frac{\varepsilon^2_{\text{STO}}}{\varepsilon^2_{\text{GaAs}}}$ under a similar scattering geometry, and $\frac{\varepsilon^2_{\text{STO}}}{\varepsilon^2_{\text{GaAs}}}\approx\frac{\Xi_{\text{GaAs}}^2}{\Xi_{\text{STO}}^2}$ if $\sigma_e$ is assumed to be similar in the two materials. Such assumption is not unreasonable because the heaviest elements in these materials are not far off in the atomic number and we are not hitting X-ray resonance in between their edges.

Now we consider thermoelastic coupling as the electron-lattice coupling mechanism. 
The temperature rise caused by the absorption of one X-ray photon is $\Delta T=\frac{h\nu}{CN/N_A}$ where $C$ is the heat capacity in J/(K$\cdot$ mol), $N_A$ is the Avocadaro number. Due to thermal expansion, the strain caused by temperature rise is $\varepsilon=\alpha \Delta T$, where $\alpha$ is the thermal expansion coefficient. Therefore $\frac{\varepsilon}{h\nu}=\frac{\alpha N_A V}{C\Omega}=\frac{3\alpha N_A V}{4\pi C\sigma_e^3}$. To compare $\varepsilon$ in the two materials we only need to compare their $\alpha/C$.
$\frac{[C(\mathbf{Q},t)/S_0(\mathbf{Q})]_{\text{STO}}}{[C(\mathbf{Q},t)/S_0(\mathbf{Q})]_{\text{GaAs}}}\approx \frac{\varepsilon^2_{\text{STO}}}{\varepsilon^2_{\text{GaAs}}}=\frac{(\alpha/C)^2_{\text{STO}}}{(\alpha/C)^2_{\text{GaAs}}}$ 
For SrTiO$_3$ , $\alpha=3.23\times 10^{-5}$K$ ^{-1}$,
$C=98 $J/(K$\cdot$ mol)~\cite{PhysRevB.53.3013}. For GaAs, $\alpha=6\times 10^{-6} $K$^{-1}$, 
$C=45$ J/(K$\cdot$ mol)~\cite{glazov2000thermal}.
This results in only a factor of 6 larger signal in SrTiO$_3$.

\clearpage
\bibliography{Citations_for_HXRSND_paper,citations_phsun,Citations_for_HXRSND_paper_SWT_update, citationsReis}

\end{document}